\documentclass[journal]{IEEEtran}

\usepackage{cite}
\usepackage{amsmath,amssymb,amsfonts}
\usepackage{algorithmic}
\usepackage{array}
\usepackage[caption=false,font=footnotesize]{subfig}
\usepackage{stfloats}
\usepackage{url}
\usepackage{mathrsfs}
\usepackage[dvips]{graphicx}

\newcommand{\diag}{\mathrm{diag}}
\newcommand{\ve}{\mathrm{vec}}
\newcommand{\T}{\mathrm{T}}
\newcommand{\He}{\mathrm{H}}
\newcommand{\jim}{\jmath}

\newenvironment{mat}[1]{\left[\begin{array}{#1}}{\end{array}\right]}
\newcommand{\bmx}[1]{\begin{mat}{#1}}
\newcommand{\emx}{\end{mat}}

\newcommand{\gssa}[4]{\mbox{\boldmath $#1$}_{#2}^{#3}(#4)}
\newcommand{\ga}[2]{\mbox{\boldmath $#1$}(#2)}
\newcommand{\gss}[3]{\mbox{\boldmath $#1$}_{#2}^{#3}}
\newcommand{\g}[1]{\mbox{\boldmath $#1$}}

\newcommand{\sbm}[1]{\mbox{\scriptsize $\g{#1}$}}
\newcommand{\sbms}[2]{\mbox{\scriptsize $\gss{#1}{#2}{}$}}

\newcommand{\pref}[1]{(\ref{#1})}

\begin{document}

\title{A Tensor-based Approach to Joint Channel Estimation / Data Detection in Flexible Multicarrier MIMO Systems}

\author{Eleftherios Kofidis
\thanks{E. Kofidis is with the Department of Statistics and Insurance Science, University of Piraeus, Piraeus, 185~34 Greece (e-mail: kofidis@unipi.gr) and the Computer Technology Institute \& Press ``Diophantus" (CTI), Patras, 26~504 Greece.}
}

\maketitle

\begin{abstract}
Filter bank-based multicarrier (FBMC) systems have attracted increasing attention recently in view of their many advantages over the classical cyclic prefix (CP)-based orthogonal frequency division multiplexing (CP-OFDM) modulation. However, their more advanced structure (resulting in, for example, self interference) complicates signal processing tasks at the receiver, including synchronization, channel estimation and equalization. In a multiple-input multiple-output (MIMO) configuration, the multi-antenna interference has also to be taken into account. (Semi-) blind receivers, of increasing interest in (massive) MIMO systems, have been little studied so far for FBMC and mainly for the single-antenna case only. The design of such receivers for \emph{flexible} MIMO FBMC systems, unifying a number of existing FBMC schemes, is considered in this paper through a tensor-based approach, which is shown to encompass existing joint channel estimation and data detection approaches as special cases, adding to their understanding and paving the way to further developments. Simulation-based results are included, for realistic transmission models, demonstrating the estimation and detection performance gains from the adoption of these receivers over their training only-based counterparts.
\end{abstract}



\section{Introduction}
\label{sec:intro}

\IEEEPARstart{F}{ilter} bank-based multicarrier (FBMC) systems~\cite{f11} have recently attracted increasing attention as a competent alternative to the classical cyclic prefix (CP)-based orthogonal frequency division multiplexing (CP-OFDM) modulation, in a wide range of applications, including both wireless (e.g., ad-hoc public safety) and wired (e.g., fiber optics) communications~\cite{rmkb17}. The potential of FBMC transmission stems from its increased ability to carrying a flexible spectrum shaping, along with a major increase in spectral efficiency and robustness to synchronization requirements, features of fundamental importance in modern and future networks. 
However, its more advanced structure (resulting in, for example, self interference) complicates signal processing tasks at the receiver, including synchronization, channel estimation and equalization~\cite{rmkb17}. In a multiple-input multiple-output (MIMO) configuration, the multi-antenna interference has also to be taken into account, which can be a challenge in, for example, offset quadrature amplitude modulation (OQAM)-based FBMC~\cite{pczlkhmc16}. Nevertheless FBMC(/OQAM) is being also considered for massive MIMO systems~\cite{rmhl18,b18} and has also shown its potential in the envisaged millimeter wave (mmWave) communications~\cite{nzlcr17}.

Although FBMC research has been rapidly advancing in the last decade or so, resulting in a number of well performing techniques for receiver design, blind (i.e., using no pilots) or semi-blind (i.e., with limited pilot information) FBMC methods have been very little studied so far, and mainly for the single-antenna case (see~\cite{kbmbs17} and~\cite{bcrylmbs17} for a recent review in (semi-) blind FBMC/OQAM channel estimation and equalization, respectively). With the advent of massive MIMO systems and the difficulties they imply in training-based estimation, semi-blind or joint channel estimation/data detection (JCD) techniques\footnote{JCD for FBMC/OQAM has been also recently considered as an application of the modern neural network trend~\cite{llzl18}.} are being considered as an alternative approach~\cite{rkpt10,pvb17,cgs18}, re-surging the interest in blind source separation (BSS) for wireless communications~\cite{zlz18}. Interestingly, (semi-)blind MIMO techniques have been also recently considered in a couple of works as a potential solution to the pilot contamination problem in massive MIMO-FBMC/OQAM-based configurations (see~\cite{mfmfcf15,krlmglz17} and references therein). A recent performance analysis for MIMO-OFDM~\cite{lmab17} showed that the attainable reduction in pilot size through a semi-blind approach can exceed 95\%. 

Tensor models and methods have been extensively studied for communications applications~\cite{a07}, including system modeling and receiver design for single-input multiple-output (SIMO) and MIMO systems, both in a general~\cite{a07,rg18,qfsy18,kks18,ramh19} and a multicarrier and/or spread spectrum~\cite{sb01,js03,js04,rcs06,nl08,sh10a,zgx10,xfd11,lcas12,xcj12,gl13,lcsa13,sd13,loz13,rrc15,wzj16,ve17,nha17,nchve17,nha18,aacs19,aah19} setup. The inherent ability of tensor models to capture the relations among the various system's dimensions, in a way that is \emph{unique} under mild conditions and/or constraints, has been exploited in problems of jointly estimating synchronization parameters, channel(s), and transmitted data symbols.  
Tensorial approaches have proven their unique advantages not only in their `natural' applications in (semi-)blind receivers~\cite{sb01,js03,nl08,r15,rrc15} but also in the design of training-based high performance receivers for challenging scenarios (for example, in MIMO relaying systems~\cite{rkx12}). Notably, in OFDM applications~\cite{sb01,js03,gl13}, performance close to that with perfect knowledge of the system parameters has been achieved~\cite{js03}. 

In the light of their successful application in OFDM (semi-) blind estimation problems, tensor-based techniques are considered here in the context of multi-antenna flexible FBMC systems. The so-called \emph{flexible} FBMC system presented in~\cite{gls14} encompasses, in a unifying manner, a number of FBMC schemes, including filtered multi-tone (FMT), FBMC/OQAM, and CP-OFDM among others.\footnote{See~\cite{mkk16} for a more extensive unification effort.} The problem of JCD in this context is studied in this paper through a tensorial approach. 
The difficulty comes mainly from the self-interference effect, which challenges the receiver design even under the commonly made simplifying assumption of channels of low selectivity, also adopted in this paper. The tensor decomposition algorithm developed is shown to cover, as special cases, existing iterative procedures for joint parameter and data estimation in MIMO (multicarrier) systems (e.g., \cite{w13,smjv19}). Simulation-based results are included, for realistic transmission models, demonstrating the estimation and detection performance gains from the adoption of these receivers over their training only-based counterparts.
This paper is an extension of the earlier related work on FBMC/OQAM~\cite{kca16,kca17} and its generalization to the flexible multicarrier context. 

The rest of the paper is organized as follows. Section~\ref{sec:fFBMC} describes the flexible FBMC system, along with example cases, and explains the self-interference generation mechanism. Details for the latter are provided in Appendix~\ref{sec:TF}. The system model considered is described in Section~\ref{sec:SM} and its tensorial formulation is discussed in Section~\ref{sec:tensor}. More details are given in Appendix~\ref{sec:PARATUCK2}. The semi-blind JCD algorithm is developed in Section~\ref{sec:SB}, where it is shown to reduce to existing techniques under simplifying assumptions. The implications of relaxing the assumption of temporally white noise are detailed in Appendix~\ref{sec:noise}. Section~\ref{sec:sims} presents and discusses the simulation results and Section~\ref{sec:concls} concludes the paper. 

\noindent 
\emph{Notation}:
Vectors, matrices and higher-order tensors are denoted by boldface lowercase, uppercase, and calligraphic uppercase letters, respectively. The superscripts $^{\T}$, $^{\He}$, $^{\dag}$, and $^{*}$ respectively stand for transposition, Hermitian transposition, Moore-Penrose pseudoinverse, and complex conjugation. 
The symbols $\otimes$, $\diamond$, and $\odot$ denote (left) Kronecker, Khatri-Rao, and Hadamard product, respectively. The operator $\ve$ stacks the columns of a matrix into a column vector. The reverse operation is performed by $\mathrm{unvec}$. If $\g{a}$ is a vector, $\diag(\g{a})$ is the diagonal matrix with $\g{a}$ on its main diagonal. $\gss{I}{R}{}$ is the identity matrix of order $R$, $\gss{1}{R}{}$ stands for the $R\times 1$ vector of all ones, and $\gss{0}{Q\times R}{}$ denotes an all zeros matrix of dimensions $Q\times R$ (or  understood from the context if not made explicit). $\gss{\mathcal{I}}{d,P}{}$ stands for the order-$d$ $P\times P\times\cdots\times P$ identity tensor. The $n$th mode unfolding (or matricization) of a higher-order tensor $\g{\mathcal{A}}$ is denoted by $\gss{A}{(n)}{}$. The mode-$n$ product of a tensor with a matrix, say $\g{\mathcal{A}}\times_{n}\g{B}$, yields a tensor with mode-$n$ unfolding $\g{B}\gss{A}{(n)}{}$. The Frobenius norm of the tensor $\g{\mathcal{A}}$ is denoted by $\|\g{\mathcal{A}}\|_{\mathrm{F}}$. The $(i,j)$th entry of a matrix $\g{A}$ is denoted by $[\g{A}]_{i,j}$. The Matlab colon notation is adopted to denote parts of an array; for example, the row vector $\ga{A}{i,:}$ is the $i$th row of the matrix $\g{A}$ and the column vector $\ga{A}{:,j}$ is its $j$th column. The real and imaginary parts of a complex quantity are denoted by $\Re\{\cdot\}$ and $\Im\{\cdot\}$, respectively, and $E\{\cdot\}$ is the expectation operator. $\jim=\sqrt{-1}$ is the imaginary unit.

\section{Flexible FBMC}
\label{sec:fFBMC}

The output of the synthesis filter bank (SFB) is given by
\begin{equation}
s(l)=\sum_{m=0}^{M-1}\sum_{n=0}^{N-1}x_{m,n}g_{m,n}(l), l=0,1,\ldots,L-1,
\label{eq:s}
\end{equation}
where $M$ and $N$ are the (even) number of subcarriers and the total number of FBMC symbols transmitted, respectively,
$x_{m,n}\stackrel{\triangle}{=}d_{m,n}e^{\jim \phi_{m,n}}$ is the (phase rotated) symbol transmitted at the frequency-time (FT) point $(m,n)$, and
\begin{equation}
g_{m,n}(l)\stackrel{\triangle}{=}g(l-nM_{\mathrm{ss}})e^{\jim \frac{2\pi}{P}ml},
\label{eq:gmn}
\end{equation}
with $g$ being the prototype filter impulse response, assumed real symmetric of length $L_{g}$ and unit energy, and $M_{\mathrm{ss}}$ and $P$ respectively denoting the symbol period and the subcarrier period (in samples).\footnote{ss in $M_{\mathrm{ss}}$ stands for \emph{s}amples per \emph{s}ymbol.} Note that $o\stackrel{\triangle}{=}\frac{P}{M}\geq 1$ is the oversampling factor~\cite{zlcp13} and $P-M$ can be taken as the number of virtual (non-active) subcarriers~\cite{gls14}. The length of the FBMC signal is $L=L_{g}+(N-1)M_{\mathrm{ss}}$. The above formulation unifies a number of existing modulation schemes, including CP-OFDM, FBMC/OQAM, FMT, etc.~\cite[Table~I]{lgss13}. For example, in FBMC/OQAM, $P=M$, $M_{\mathrm{ss}}=\frac{M}{2}$, $d_{m,n}$ are real (pulse amplitude modulated (PAM)) symbols, $\phi_{m,n}$ is such that $\phi_{m,n}\mod \pi=(m+n)\frac{\pi}{2}$, and $L_{g}=KM$ (or $L_g=KM+1$~\cite{b01}), with $K$ being the overlapping factor~\cite{rmkb17}. 
CP-OFDM, with a CP of length $M_{\mathrm{CP}}$, results with $P=M$, $M_{\mathrm{ss}}=M+M_{\mathrm{CP}}$, and $g$ being a rectangular pulse of length $L_g=M_{\mathrm{ss}}$ and amplitude $\frac{1}{\sqrt{M}}$ while the CP insertion is represented through an appropriate definition of the phase rotation~\cite{lgss13}, which can be easily verified to be $\phi_{m,n}=-\frac{2\pi}{M}m(n+1)M_{\mathrm{CP}}$.
The Generalized Frequency Division Multiplexing (GFDM) scheme~\cite[Chapter~1]{rmkb17} can be so described as well, with a circular instead of a linear time shifting in~\pref{eq:gmn} and, again, an appropriate phase rotation representing the CP. 

The corresponding (in a back-to-back configuration) output of the analysis filter bank (AFB) at the frequency-time point $(p,q)$ can be written as
\begin{equation}
y_{p,q}=\sum_{l=0}^{L-1}s(l)\tilde{g}_{p,q}^{*}(l),
\label{eq:ypq}
\end{equation}
where the receive prototype filter, $\tilde{g}$, coincides with the transmit one, $g$, in FBMC schemes free from guard intervals such as CP, whereas its definition incorporates the guard interval removal operation whenever required, as in, for example, CP-OFDM:\footnote{In this sense, CP-OFDM and other modulation schemes of this kind could be categorized as \emph{bi-}orthogonal.} 
\begin{equation}
\tilde{g}(l)=\left\{\begin{array}{cc} g(l), & M_{\mathrm{CP}}\leq l \leq M+M_{\mathrm{CP}}-1 \\ 0, & 0\leq l \leq M_{\mathrm{CP}}-1 \end{array}\right.
\label{eq:gtilde}
\end{equation}
Using~\pref{eq:s} in~\pref{eq:ypq} results in
\begin{eqnarray}
y_{p,q} & = & \sum_{l=0}^{L-1}\sum_{m=0}^{M-1}\sum_{n=0}^{N-1}x_{m,n}g_{m,n}(l)\tilde{g}_{p,q}^{*}(l) \nonumber \\
& = & \sum_{m=0}^{M-1}\sum_{n=0}^{N-1}x_{m,n}I_{m,n}^{p,q} 
\label{eq:ypqI}
\end{eqnarray}
with 
\begin{equation}
I_{m,n}^{p,q}\triangleq \sum_{l}g_{m,n}(l)\tilde{g}^{*}_{p,q}(l)
\label{eq:Imnpq}
\end{equation}
being the inter-symbol (ISI) and inter-carrier (ICI) interference weights, with $I_{p,q}^{p,q}=1$. Substituting for $g_{m,n}(l)$ in~\pref{eq:Imnpq} and similarly for $\tilde{g}_{p,q}(l)$ yields
\begin{eqnarray*}
I_{m,n}^{p,q} \!\!\!\! & = & \!\!\!\! \sum_{l}g(l-nM_{\mathrm{ss}})\tilde{g}(l-qM_{\mathrm{ss}})e^{\jim\frac{2\pi}{P}(m-p)l} \\
\!\!\!\! & = & \!\!\!\! e^{\jim\frac{2\pi}{P}(m-p)qM_{\mathrm{ss}}}\sum_{l=l_0}^{l_1}g(l-(n-q)M_{\mathrm{ss}})\tilde{g}(l)e^{\jim\frac{2\pi}{P}(m-p)l}
\end{eqnarray*}
with $l_0\triangleq \max(0,(n-q)M_{\mathrm{ss}}),l_1\triangleq \min(L_g-1,L_g-1+(n-q)M_{\mathrm{ss}})$, whereby $I_{m,n}^{p,q}$ is seen to depend on the subcarrier and time indexes through their \emph{differences} $u\triangleq m-p$ and $v\triangleq n-q$ as well as on the time position of the FT point considered, $q$. It will thus henceforth be also denoted as $I_{u,v}$ with the understanding that it also depends on $q$:
\begin{equation}
I_{u,v} = \varepsilon^{uq}\sum_{l=l_0}^{l_1}g(l-vM_{\mathrm{ss}})\tilde{g}(l)e^{\jim\frac{2\pi}{P}ul},
\label{eq:Iuv}
\end{equation}
where $\varepsilon \triangleq e^{\jim\frac{2\pi}{P}M_{\mathrm{ss}}}$. In most of the FBMC schemes (except for the OFDM-based ones), $P=M_{\mathrm{ss}}$ or $P=2M_{\mathrm{ss}}$~\cite[Table~I]{lgss13} in which cases $\varepsilon$ becomes~1 or~$-1$, respectively. It is easy to see that $I_{u,v}=0, (u,v)\neq (0,0)$ for the schemes with rectangular pulses of length $M_{\mathrm{ss}}$ such as CP-OFDM. On the other hand, for \emph{intrinsically} filter bank-based modulations, there is non-negligible self interference, which can often be assumed to be confined in the first-order time-frequency neighborhood of the FT point under consideration (see, for example, \cite{l08} for the FBMC/OQAM case) based on the assumption of prototype filter designs with good time-frequency localization. 
Appendix~\ref{sec:TF} shows that the interference weights in this $3\times 3$ neighborhood follow the pattern below:
\begin{equation}
\begin{array}{ccc}
\Delta_q^* & B_q^* & \Delta_{q+1}^* \\
& & \\
\Gamma & 1 & \Gamma \\
& & \\
\Delta_q & B_q & \Delta_{q+1}
\end{array}
\label{eq:pattern}
\end{equation}
where the horizontal direction corresponds to time and the vertical one to frequency and the involved quantities are given by
\begin{eqnarray*}
B_q & = & \varepsilon^q \sum_{l=0}^{L_{g}-1}g(l)\tilde{g}(l)e^{\jim \frac{2\pi}{P}l}, \\
\Gamma & = & \sum_{l=0}^{L_{g}-1-M_{\mathrm{ss}}}g(l+M_{\mathrm{ss}})\tilde{g}(l), \\
\Delta_q & = & \varepsilon^q\sum_{l=0}^{L_{g}-1-M_{\mathrm{ss}}}g(l+M_{\mathrm{ss}})\tilde{g}(l)e^{\jim\frac{2\pi}{P}l}.
\end{eqnarray*}
The reader is referred to Appendix~\ref{sec:TF} for more detailed expressions and a discussion of important special cases.
The interference pattern (also incorporating the phase rotations) for FBMC/OQAM was first derived in~\cite{kkrt13}.

In summary, if $\Omega_{p,q}\triangleq \{(m,n): |m-p|\leq 1 \mbox{\ and\ } |n-q|\leq 1\}$ is the time-frequency neighborhood above, the response of the flexible FBMC transmultiplexer at the FT point $(p,q)$ to a symbol transmitted at the same FT point can be approximately written as 
\begin{equation}
y_{p,q}=x_{p,q}+\sum_{(m,n)\in\Omega^{\prime}_{p,q}}I_{m,n}^{p,q}x_{m,n}
\label{eq:ypqOmega}
\end{equation}
with $\Omega^{\prime}_{p,q}\triangleq \Omega_{p,q}\backslash (p,q)$ and the $I_{m,n}^{p,q}$'s being given by~\pref{eq:pattern}.

\section{System Model}
\label{sec:SM}

\subsection{Single-Input Single-Output System}
\label{subsec:SISO}

Let the FBMC signal pass through a channel with impulse response (CIR) $h$ of length $L_h$, which is assumed to be invariant over the duration of the transmitted frame of $L$ samples. 
Assuming also the presence of a carrier frequency offset (CFO), $\nu$, normalized to the subcarrier spacing\footnote{$\nu$ can be taken as the \emph{residual} CFO (after a coarse frequency synchronization has been achieved), not exceeding (in absolute value) 0.5~\cite{js03}, \cite[Chapter~10]{rmkb17}.}, the noisy channel output can be written as 
\begin{equation}
r(l)=e^{\jim \frac{2\pi}{P}\nu l} \sum_{k=0}^{L_h-1}h(k)s(l-k) + \eta(l),
\label{eq:r}
\end{equation}
where the noise $\eta(l)$ is white Gaussian with zero mean and variance $\sigma^2$. A timing offset and a carrier phase error could also be considered, as part of the complex channel model~\cite{lgss13}. 

For the common case of a relatively (to the FBMC symbol duration) low channel delay spread, one can use an argument analogous to that used in~\cite{l08} for FBMC/OQAM to arrive at the following approximation of~\pref{eq:r}:
\begin{equation}
r(l)\approx e^{\jim \frac{2\pi}{P}\nu l}\sum_{n=0}^{N-1}\sum_{m=0}^{M-1}H_{m}x_{m,n}g_{m,n}(l) + \eta(l),
\label{eq:rapprox}
\end{equation}
where $H_{m}$, $m=0,1,\ldots,P-1$ is the $P$-point channel frequency response (CFR), i.e., $H_m = \sum_{l=0}^{L_h-1}h(l)e^{-\jim \frac{2\pi}{P}lm}$. 
The AFB output at the FT point $(p,q)$ is then given by
\begin{equation}
y_{p,q}=\sum_{l}r(l)\tilde{g}_{p,q}^{*}(l) + w_{p,q},
\label{eq:ypqr}
\end{equation}
where
\begin{equation}
w_{p,q}=\sum_{l}\eta(l)\tilde{g}^{*}_{p,q}(l)
\label{eq:wpq}
\end{equation}
is the frequency-domain noise, and, using~\pref{eq:rapprox},
\[
y_{p,q} \approx \sum_{l}e^{\jim \frac{2\pi}{P}\nu l}\sum_{n=0}^{N-1}\sum_{m=0}^{M-1}H_{m}x_{m,n}g_{m,n}(l)\tilde{g}_{p,q}^{*}(l) + w_{p,q} 
\] 
or equivalently
\begin{equation}
y_{p,q} \approx \sum_{n=0}^{N-1}e^{\jim \frac{2\pi}{P}\nu n M_{\mathrm{ss}}} \sum_{m=0}^{M-1}H_{m}x_{m,n}\sum_{l}g^{\prime}_{m,n}(l)
\tilde{g}^{*}_{p,q}(l) + w_{p,q},
\label{eq:ypqapprox}
\end{equation}
with $g^{\prime}_{m,n}(l)$ defined as in~\pref{eq:gmn} but with $g(l)$ replaced by $g^{\prime}(l)=g(l)e^{\jim \frac{2\pi}{P}\nu l}$ (see \cite[Section~4.1]{lgs14}, \cite[eqs.~(24)--(25)]{mtb15} for the FBMC/OQAM case). If the CFO is assumed to be small enough so that $\frac{2\pi}{P}|\nu| (L_g-1)\ll 1$, i.e., the incremental phase shift within the support of $g$ due to the frequency offset can be assumed to be negligible~\cite{clr07}, then \pref{eq:ypqapprox} becomes
\begin{equation}
y_{p,q}\approx \sum_{n=0}^{N-1}e^{\jim \frac{2\pi}{P}\nu n M_{\mathrm{ss}}} \sum_{m=0}^{M-1}H_{m}x_{m,n}I_{m,n}^{p,q}+w_{p,q}.
\label{eq:ypqnu}
\end{equation}
When, in addition, no ICI is present (as in, e.g., FMT), one can write
\[
y_{p,q}\approx \sum_{n=0}^{N-1} e^{\jim \frac{2\pi}{P}\nu n M_{\mathrm{ss}}} H_{p}x_{p,n}I_{p,n}^{p,q}+w_{p,q}
\]
and in the absence of ISI as well (e.g., in CP-OFDM)~\cite{lrc10}
\begin{equation}
y_{p,q}\approx e^{\jim \frac{2\pi}{P}\nu q M_{\mathrm{ss}}} H_{p}x_{p,q}+w_{p,q}.
\label{eq:io}
\end{equation}

Given the assumption leading to~\pref{eq:ypqOmega} and assuming the channel's frequency selectivity to be low enough so that its CFR is invariant over $\Omega_{p,q}$, one arrives at the following simplified input-output model
\begin{equation}
y_{p,q}\approx H_{p}b_{p,q}+w_{p,q},
\label{eq:ypqlow}
\end{equation}
where
\begin{equation}
b_{p,q}=e^{\jim \frac{2\pi}{P}\nu q M_{\mathrm{ss}}}x_{p,q}+\sum_{(m,n)\in\Omega^{\prime}_{p,q}}e^{\jim \frac{2\pi}{P}\nu n M_{\mathrm{ss}}}x_{m,n}I_{m,n}^{p,q}
\label{eq:cpq}
\end{equation}
can be called the \emph{virtual} symbol transmitted at $(p,q)$. This term and the model~\pref{eq:ypqlow} are quite common in the FBMC/OQAM literature~\cite{rmkb17} as they represent the translation of the problem into a form similar to that of the conventional CP-OFDM.

Writing~\pref{eq:cpq} in the form
\[
b_{p,q} = e^{\jim \frac{2\pi}{P}\nu q M_{\mathrm{ss}}}\left(x_{p,q}+\sum_{(m,n)\in\Omega^{\prime}_{p,q}}e^{\jim \frac{2\pi}{P}\nu (n-q) M_{\mathrm{ss}}}x_{m,n}I_{m,n}^{p,q}\right)
\]
and observing that the small CFO assumption along with the fact that $|n-q|$ is only~1 imply that $e^{\jim \frac{2\pi}{P}\nu (n-q) M_{\mathrm{ss}}}\approx 1$ leads to the following simplification of~\pref{eq:ypqlow} (in the form of~\pref{eq:io}):
\begin{equation}
y_{p,q}\approx e^{\jim \frac{2\pi}{P}\nu q M_{\mathrm{ss}}}H_{p}c_{p,q}+w_{p,q},
\label{eq:ypqclow}
\end{equation}
with $c_{p,q}$ denoting the CFO-free virtual symbol (shown also in~\pref{eq:ypqOmega}). 

\subsection{MIMO System}
\label{subsec:MIMO}

Consider now the extension of the above model to a MIMO configuration with $N_{\mathrm{T}}$ transmit and $N_{\mathrm{R}}$ receive antennas. Let
\begin{equation}
\gss{h}{}{(r,t)}=\bmx{cccc} H^{(r,t)}_{0} & H^{(r,t)}_{1} & \cdots & H^{(r,t)}_{M-1}\emx^{\T}
\label{eq:Hrt}
\end{equation}
be the frequency response of the channel from the transmit (Tx) antenna $t$ to the receive (Rx) antenna $r$ and denote by $x_{m,n}^{(t)}$ the symbol transmitted at the FT point $(m,n)$ from the $t$th Tx antenna, with $c^{(t)}_{m,n}$ being the corresponding virtual symbol. Eq.~\pref{eq:ypqclow} for the $r$th Rx antenna then writes as follows:
\begin{equation}
y_{p,q}^{(r)}\approx\sum_{t=1}^{N_{\T}} e^{\jim \frac{2\pi}{P}\nu q M_{\mathrm{ss}}}H_{p}^{(r,t)}c^{(t)}_{p,q}+w^{(r)}_{p,q},
\label{eq:ypqclowMIMO}
\end{equation}
where, as usual, the noise signals at different Rx antennas, $w_{p,q}^{(r)}$, are uncorrelated with each other and the same CFO, $\nu$, is assumed to be present for each pair of Tx-Rx antennas. It should be noted that this is a plausible assumption in single-user MIMO systems~\cite{yg05} (but not realistic in a multi-user setup~\cite{yg05,cwmn08}). Collecting the samples received at the $r$th antenna into an $M\times N$ matrix yields
\begin{equation}
\gss{Y}{}{(r)}\triangleq [y_{m,n}^{(r)}]=\sum_{t=1}^{N_{\T}}\diag(\gss{h}{}{(r,t)})\gss{C}{}{(t)}\diag(\gss{e}{\nu}{})+\gss{W}{}{(r)},
\label{eq:Yr}
\end{equation}
where $\gss{W}{}{(r)}$ is similarly built from $w_{m,n}^{(r)}$, the $M\times N$ matrix $\gss{C}{}{(t)}\triangleq [c_{m,n}^{(t)}]$ collects the virtual symbols for Tx antenna $t$, and the $N\times 1$ vector $\gss{e}{\nu}{}$ contains the CFO-induced factors $e^{\jim \frac{2\pi}{P}\nu n M_{\mathrm{ss}}}$, $n=0,1,\ldots,N-1$. 

\section{Tensor-based Formulation}
\label{sec:tensor}

Eq.~\pref{eq:Yr} can be also written as
\begin{eqnarray*}
\gss{Y}{}{(r)} & = & \bmx{ccc} \diag(\gss{h}{}{(r,1)}) & \cdots & \diag(\gss{h}{}{(r,N_{\T})}) \emx \times 
\\ 
& & \underbrace{\bmx{c} \gss{C}{}{(1)} \\ \gss{C}{}{(2)} \\ \vdots \\ \gss{C}{}{(N_{\T})} \emx}_{\sbm{C}}
\diag(\gss{e}{\nu}{})+\gss{W}{}{(r)},
\end{eqnarray*}
or equivalently
\begin{equation}
\gss{Y}{}{(r)} \! = \! \underbrace{(\gss{1}{N_{\T}}{\T}\otimes \gss{I}{M}{})}_{\sbm{\Gamma}}\,\diag(\gss{h}{}{(r,\cdot)})\,
\underbrace{\g{C} \, \diag(\gss{e}{\nu}{})}_{\sbm{C}_{\nu}} + \gss{W}{}{(r)}, \label{eq:Yrnu}
\end{equation}
with
\[
\gss{h}{}{(r,\cdot)}\triangleq \bmx{cccc} \gss{h}{}{(r,1)\T} & \gss{h}{}{(r,2)\T} & \cdots & \gss{h}{}{(r,N_{\T})\T}\emx^{\T}.
\]
Viewing the above as the $r$th frontal slice of an $M\times N\times N_{\mathrm{R}}$ tensor $\g{\mathcal{Y}}$ leads to the conclusion that its noise-free part, $\g{\bar{\mathcal{Y}}}$, obeys a \emph{Canonical Polyadic Decomposition (CPD) (or PARAllel FACtor analysis (PARAFAC))} model~\cite{sdfhpf17} of rank $N_{\T}M$, namely
\begin{equation}
\g{\mathcal{Y}}=\underbrace{\gss{\mathcal{I}}{3,N_{\T}M}{}\times_1 \g{\Gamma} \times_2 \gss{C}{\nu}{\T} \times_3 \g{H}}_{\sbm{\bar{\mathcal{Y}}}} + \g{\mathcal{W}},
\label{eq:CPD}
\end{equation}
where $\g{\mathcal{W}}$ has the $\gss{W}{}{(r)}$'s at its frontal slices and the $N_{\mathrm{R}}\times N_{\T}M$ matrix $\g{H}$ has rows $\gss{h}{}{(r,\cdot)\T}$, $r=1,2,\ldots,N_{\mathrm{R}}$. Observe that the virtual symbols and the CFO are mixed in the $\gss{C}{\nu}{}$ factor above, which prevents their estimation\footnote{Unless additional information (e.g., appropriate training) and/or estimation methods (e.g., \cite[Chapter~10]{rmkb17}, \cite{cwmn08,sh10b,sh17,fzwlx11,glm14}) are employed.} from a direct decomposition of the tensor (see Appendix~\ref{sec:PARATUCK2} for more about the tensorial formulation). Thus, perfect frequency synchronization (i.e., zero CFO) will be assumed in the sequel (hence $\gss{C}{\nu}{}=\g{C}$), for the sake of simplicity. 

The \emph{uniqueness} property of the CPD, which can be trusted to hold under \emph{mild} conditions~\cite{sdfhpf17}, can be taken advantage of in such a setup (in a way analogous to that followed in OFDM; see, e.g., \cite{js03}, etc.) to \emph{blindly} estimate the unknown factors, $\g{H}$ and $\g{C}$, from the tensor $\g{\mathcal{Y}}$. The inherent scaling ambiguity can be overcome through, for example, the transmission of limited training information (as in, e.g., \cite{js03}), as explained in detail in Section~\ref{sec:sims}. The \emph{a-priori} knowledge of one of the factors (namely $\g{\Gamma}$) in the CPD above trivially resolves the permutation ambiguity. 

Kruskal's condition, the most well-known and general sufficient condition for the uniqueness of a CPD, would be stated for~\pref{eq:CPD} as follows~\cite{sdfhpf17}:
\[
k_{\sbm{\Gamma}}+k_{\sbm{C}^{\T}}+k_{\sbm{H}}\geq 2N_{\T}M+2,
\]
where, for a given matrix $\g{A}$, $k_{\sbm{A}}$ denotes its Kruskal or k-rank, that is, the largest number $k$ such that any subset of $k$ columns of this matrix is linearly independent. However, since $\g{\Gamma}$ is \emph{a-priori} known, other, more suitable for this case, uniqueness conditions should be considered, as given in~\cite{sd15}. Thus, for a SIMO system, where $\g{\Gamma}=\gss{I}{M}{}$ is of full column rank (f.c.r.), \cite[Proposition~3.1]{sd15} applies, which guarantees uniqueness of the CPD in this case. Indeed, the transpose of the mode-1 unfolding of $\g{\bar{\mathcal{Y}}}$ can be written from~\pref{eq:CPD} as
\begin{equation}
\gss{\bar{Y}}{(1)}{\T}=(\g{H}\diamond \gss{C}{}{\T})\gss{\Gamma}{}{\T},
\label{eq:Y1T}
\end{equation}
or
\begin{equation}
\gss{\bar{Y}}{(1)}{\T}=\g{H}\diamond \gss{C}{}{\T},
\label{eq:Y1TSIMO}
\end{equation}
whereby uniqueness follows from the fact that the $j$th column of $\gss{\bar{Y}}{(1)}{\T}$ is the vectorized version of the rank-1 matrix $\ga{C}{j,:}^{\T}\ga{H}{:,j}^{\T}$ and hence $\g{H}$ and $\gss{C}{}{\T}$ can be computed by the rank-1 approximations of the columns of $\gss{\bar{Y}}{(1)}{\T}$. Of course, to avoid having trivial rank-1 approximation problems, the condition of $N_{\mathrm{R}}\geq 2$, that is, of a non-trivial spatial dimension, is required.
For a non f.c.r. $\g{\Gamma}$, namely when $N_{\T}>1$, Proposition~3.2 of~\cite{sd15} applies, which however can be verified not to guarantee uniqueness of the above CPD (since in that case $k_{\sbm{\Gamma}}=1$). Nevertheless, there is a way to get past this, based on the observation that $\gss{\bar{Y}}{(1)}{\T}$ in~\pref{eq:Y1T} is a sum of Khatri-Rao products, and is discussed in the next section.

In the presence of noise, the CPD in~\pref{eq:CPD} can only be \emph{approximated}. Given the fact that the noise at the AFB output is colored Gaussian, the maximum likelihood (ML) JCD problem becomes a \emph{weighted} least squares (LS) one. The correlation in the noise tensor $\g{\mathcal{W}}$ is found in two of its three dimensions (time and frequency, not space) with corresponding covariances that can be \emph{a-priori} computed as shown in Appendix~\ref{sec:noise}. CPD estimation algorithms of the kind considered in the next section that take the noise color into account include~\cite{bss02,s04,v05,rkx12,rfmc17}. Nevertheless, since the inversion of the noise covariance is shown in Appendix~\ref{sec:noise} to be in general a daunting task, especially for large values of $M$ and $N$, and given the rareness of the use of the noise color in the FBMC estimation literature, the noise will be assumed white in this paper, for simplicity. 
The JCD problem can then be stated (in the notation of~\cite{kb09}) as 
\begin{equation}
\min_{\sbm{H},\sbm{C}}\left\|\g{\mathcal{Y}}-[\![\g{\Gamma},\gss{C}{}{\mathrm{T}},\g{H}]\!]\right\|_{\mathrm{F}}.
\label{eq:prob}
\end{equation}

\section{Semi-Blind Flexible Multicarrier MIMO Receivers}
\label{sec:SB}

Writing the (transposed) mode-1 matricization of $\g{\mathcal{Y}}$, 
\begin{eqnarray}
\gss{Y}{(1)}{\T} & = & \bmx{cccc} \gss{Y}{}{(1)} & \gss{Y}{}{(2)} & \cdots & \gss{Y}{}{(N_{\mathrm{R}})} \emx^{\T} \nonumber \\
& = & (\g{H}\diamond\gss{C}{}{\T})\gss{\Gamma}{}{\T}+\gss{W}{(1)}{\T},
\label{eq:Y1eq}
\end{eqnarray}
in the form
\[
\gss{Y}{(1)}{\T}=\sum_{t=1}^{N_{\mathrm{T}}}\bmx{c} (\gss{h}{}{(1,t)})^{\T} \\ (\gss{h}{}{(2,t)})^{\T} \\ \vdots \\ (\gss{h}{}{(N_{\mathrm{R}},t)})^{\T} \emx \diamond (\gss{C}{}{(t)})^{\T}+\gss{W}{(1)}{\T}
\]
implies that its $(m+1)$st, $m=0,1,\ldots,M-1$, column is a sum of $N_{\T}$ Kronecker products, namely
\begin{eqnarray*}
\lefteqn{\gssa{Y}{(1)}{\T}{:,m+1}=} \\ 
& & \sum_{t=1}^{N_{\T}} \bmx{c} H_{m}^{(1,t)} \\ H_{m}^{(2,t)} \\ \vdots \\ H_{m}^{(N_{\mathrm{R}},t)} \emx 
\otimes \bmx{c} c_{m,0}^{(t)} \\ c_{m,1}^{(t)} \\ \vdots \\ c_{m,N-1}^{(t)} \emx + \gssa{W}{(1)}{\T}{:,m+1},
\end{eqnarray*}
which, if reshaped to an $N\times N_{\mathrm{R}}$ matrix, will write as
\[
\underbrace{\mathrm{unvec}(\gssa{Y}{(1)}{\T}{:,m+1})}_{\sbm{Y}_{(1)}^{(m)\T}}=\gss{C}{m}{}\gss{H}{m}{\T}+
\underbrace{\mathrm{unvec}(\gssa{W}{(1)}{\T}{:,m+1})}_{\sbm{W}_{(1)}^{(m)\T}}
\]
or equivalently
\begin{equation}
\gss{Y}{(1)}{(m)}=\gss{H}{m}{}\gss{C}{m}{\T}+\gss{W}{(1)}{(m)}, \,\, m=0,1,\ldots,M-1,
\label{eq:Y1m}
\end{equation}
with the obvious definitions for the $N_{\mathrm{R}}\times N_{\T}$ matrix $\gss{H}{m}{}$ and the $N\times N_{\T}$ matrix $\gss{C}{m}{}$.
The above is recognized as a \emph{bilinear} ML BSS problem, one per subcarrier, and a number of algorithms can be invoked to solve it. For example, the so-called \emph{iterative least squares with projection (ILSP)} scheme~\cite{tvp96}, consisting of solving in a LS sense for $\gss{H}{m}{}$ given $\gss{C}{m}{}$ and conversely, in a alternating fashion, while also projecting the estimates of $\g{C}$ onto the symbols space, was proposed in~\cite{nha18} for MIMO-OFDM. Similarly with~\cite{tvp96}, one could also consider replacing projection of the LS estimate of $\gss{C}{m}{}$ by a step of enumeration over the symbols constellation, giving rise to an \emph{iterative least squares with enumeration (ILSE)} scheme, shown in~\cite{tp97} to be generally more effective than the ILSP one. The subsequent increase of complexity could be addressed with the aid of an efficient lattice search (e.g., sphere decoding~\cite{rkpt10}) procedure.

A necessary condition for the two conditional LS problems above to have a solution is that $N_{\mathrm{R}}\geq N_{\T}$ and $N\geq N_{\T}$ and can easily hold in practice. It is fundamental, for the success of the above JCD approach, that the discrete (in fact, finite) nature of the set of possible values of the $\gss{C}{m}{}$ factor be also taken into account. In view of~\pref{eq:ypqOmega} and~\pref{eq:pattern}, the virtual symbol $c^{(t)}_{m,n}$ takes values in a set of at most $Q^{\prime}\triangleq Q^9$ discrete values, where $Q$ is the order of the input constellation. This is the space that the $\gss{C}{m}{}$ estimates would be projected into or enumerated from. Identifiability conditions from~\cite{tvp96} apply only for large enough sets of \emph{independent, identically distributed (i.i.d.)} (virtual) symbols. Since the factors in~\pref{eq:Y1m} are both of full rank, the identifiability condition of~\cite{tvp96} applies, whereby it suffices for $N$ to be large enough so that $\gss{C}{m}{}$ contains all $\frac{(Q^{\prime})^{2N_{\T}}}{2}$ distinct (up to a sign) $N_{\T}$-vectors with entries belonging to a $(Q^{\prime})^2$-order constellation. The probability of non-identifiability for $N\gg \frac{(Q^{\prime})^{2N_{\T}}}{2}$ i.i.d. (multicarrier) symbols is shown in~\cite{tvp96} to approach zero exponentially fast. For large $N_{\T}$ and/or $Q^{\prime}$, the number of symbols required may become unrealistically large. More practical conditions can be found in, e.g., \cite{lpv03}, albeit only for constant modulus signals.\footnote{Thanks to Dr.~M.\ S\o rensen, formerly at KULeuven, for pointing out this paper.} A related upper bound on the probability of non-identifiability for the case of i.i.d. binary PSK input can be found in~\cite{js03}.  
The finite-valued nature of the (virtual) symbols can be also exploited through a geometry-based approach, as in~\cite{dwg18}. 

However, in general (with the exception of CP-OFDM-like schemes), the virtual symbols are not i.i.d. as they relate to the input symbols through a two-dimensional (time-frequency) filtering. Denoting the $(n+1)$st virtual multicarrier symbol by $\gss{c}{n}{(t)}\triangleq \gssa{C}{}{(t)}{:,n+1}$ and similarly for 
$\gss{x}{n}{(t)}\triangleq \gssa{X}{}{(t)}{:,n+1}$, one can use~\pref{eq:ypqOmega} and~\pref{eq:pattern} to write, for any Tx antenna $t$, 
\[
\gss{c}{n}{(t)}\!\!\!=\!\!\!\gss{x}{n}{(t)}+\gss{\mathfrak{B}}{n}{}\gss{x}{n}{(t)}+\Gamma\gss{x}{n-1}{(t)}+\Gamma\gss{x}{n+1}{(t)}+\gss{\mathfrak{D}}{n}{}\gss{x}{n-1}{(t)}+\gss{\mathfrak{D}}{n+1}{}\gss{x}{n+1}{(t)},
\]
where the $M\times M$ matrices
\[
\gss{\mathfrak{B}}{n}{}=\bmx{ccccccc} 0 & B_n & 0 & 0 & \cdots & 0 & B_n \\ B_n^* & 0 & B_n & 0 & \cdots & 0 & 0 \\ 0 & B_n^* & 0 & B_n & \cdots & 0 & 0 \\ \vdots & \ddots & \ddots & \ddots & \ddots & \ddots & \vdots \\ B_n^* & 0 & 0 & 0 & \cdots & B_n^* & 0 \emx
\]
and
\[
\gss{\mathfrak{D}}{n}{}=\bmx{ccccccc} 0 & \Delta_n & 0 & 0 & \cdots & 0 & \Delta_n \\ \Delta_n^* & 0 & \Delta_n & 0 & \cdots & 0 & 0 \\ 0 & \Delta_n^* & 0 & \Delta_n & \cdots & 0 & 0 \\ \vdots & \ddots & \ddots & \ddots & \ddots & \ddots & \vdots \\ \Delta_n^* & 0 & 0 & 0 & \cdots & \Delta_n^* & 0 \emx
\]
represent the interference in the frequency and time-frequency directions, respectively. FBMC symbol time indexes that are negative or larger than $N-1$ are taken care of by making the common assumption that the data frame is preceded and followed by inactive inter-frame gaps, resulting in negligible interference among frames~\cite{kkrt13}. Extracting the input symbols, $x_{m,n}^{(t)}$, from the virtual symbols requires self-interference cancellation in general. Nevertheless, in GFDM, for example, the modulation can be inverted (see, e.g., \cite{nchve17}) with the aid of the CP employed in this modulation scheme. In CP-OFDM (and related modulations), $\gss{C}{}{(t)}=\gss{X}{}{(t)}$ and $\gss{X}{}{(t)}=\diag([e^{-\jim\frac{2\pi}{M}mM_{\mathrm{CP}}}]_{m=0}^{M-1})\gss{D}{}{(t)}\odot\g{\mathfrak{M}}$ where $[\g{\mathfrak{M}}]_{m,n}\triangleq e^{-\jim\frac{2\pi}{M}mnM_{\mathrm{CP}}}$ and the $M\times N$ matrix $\gss{D}{}{(t)}\triangleq [d_{m,n}(t)]$ contains the original (without phase rotation) input symbols. In the FBMC/OQAM case, assuming that (as with, e.g., the PHYDYAS filter bank) $L_g=KP+1$, $B_{n}$'s and $\Delta_{n}$'s are real- and imaginary-valued, respectively, and $B_{n}=(-1)^{n}B_0$, $\Delta_{n}=(-1)^{n}\Delta_0$. Letting then
\[
\gss{Z}{Q}{}\triangleq\left[\begin{array}{cc} \gss{0}{1\times (Q-1)}{} & 1 \\ \gss{I}{Q-1}{} & \gss{0}{(Q-1)\times 1}{} \end{array}\right]
\]
denote the $Q\times Q$ matrix of circular downwards shifting and 
\[
\gss{\mathfrak{Z}}{Q}{}\triangleq\left[\begin{array}{cc} \gss{0}{1\times (Q-1)}{} & 0 \\ \gss{I}{Q-1}{} & \gss{0}{(Q-1)\times 1}{} \end{array}\right]
\]
the corresponding matrix for non-circular shifting, 
one can express $\gss{C}{}{(t)}$ in terms of $\gss{X}{}{(t)}$ as follows 
\begin{equation}
\gss{C}{}{(t)}=\gss{X}{}{(t)}+B_{0}\gss{Z}{M}{+}\gss{X}{}{(t)}\gss{S}{N}{}-\Gamma\gss{X}{}{(t)}\gss{\mathfrak{Z}}{N}{-}+\Delta_{0}\gss{Z}{M}{+}\gss{X}{}{(t)}\gss{S}{N}{}\gss{\mathfrak{Z}}{N}{+},
\label{eq:C(X)}
\end{equation}
where $\gss{S}{N}{}\triangleq\diag(1,-1,1,-1,\ldots)$ is of order $N$, 
$\gss{Z}{M}{+}$ is the circulant $M\times M$ matrix
\[
\gss{Z}{M}{+}\triangleq \gss{Z}{M}{}+\gss{Z}{M}{-1}=\left[\begin{array}{ccccccc} 0 & 1 & 0 & 0 & \cdots & 0 & 1 \\ 1 & 0 & 1 & 0 & \cdots & 0 & 0 \\ 
0 & 1 & 0 & 1 & \cdots & 0 & 0 \\ \vdots & \vdots & \ddots & \ddots & \cdots & \ddots & \vdots \\
1 & 0 & 0 & \cdots & 0 & 1 & 0 \end{array}\right],
\]
and $\gss{\mathfrak{Z}}{N}{+}$ and $\gss{\mathfrak{Z}}{N}{-}$ are similarly structured Toeplitz $N\times N$ matrices: 
\begin{eqnarray*}
\gss{\mathfrak{Z}}{N}{+}\triangleq \gss{\mathfrak{Z}}{N}{}+\gss{\mathfrak{Z}}{N}{\T} & = & \left[\begin{array}{ccccccc} 0 & 1 & 0 & 0 & \cdots & 0 & 0 \\ 1 & 0 & 1 & 0 & \cdots & 0 & 0 \\ 
0 & 1 & 0 & 1 & \cdots & 0 & 0 \\ \vdots & \vdots & \ddots & \ddots & \cdots & \ddots & \vdots \\
0 & 0 & 0 & \cdots & 0 & 1 & 0 \end{array}\right], \\
\gss{\mathfrak{Z}}{N}{-}\triangleq \gss{\mathfrak{Z}}{N}{}-\gss{\mathfrak{Z}}{N}{\T} & = & 
\left[\begin{array}{ccccccc} 0 & -1 & 0 & 0 & \cdots & 0 & 0 \\ 1 & 0 & -1 & 0 & \cdots & 0 & 0 \\ 
0 & 1 & 0 & -1 & \cdots & 0 & 0 \\ \vdots & \vdots & \ddots & \ddots & \cdots & \ddots & \vdots \\
0 & 0 & 0 & \cdots & 0 & 1 & 0 \end{array}\right].
\end{eqnarray*}
With the common definition of the phase rotation, $\phi_{m,n}=(m+n)\frac{\pi}{2}$, $\gss{X}{}{(t)}$ can be written as $\diag(\gss{\jim}{M}{})\gss{D}{}{(t)}\diag(\gss{\jim}{N}{})$ where $\gss{\jim}{M}{}\triangleq \gss{1}{M/4}{}\otimes \g{\jim}$, with 
$\g{\jim}\triangleq \bmx{cccc} 1 & \jim & -1 & -\jim \emx^{\T}$ being the 4th roots of unity, and $\gss{\jim}{N}{}$ is similarly defined. 
Substituting in~\pref{eq:C(X)} yields~\pref{eq:tildeC(X)} for the virtual symbols with phase rotation in the frequency and time directions undone, where $\gss{Z}{M}{-}\triangleq \gss{Z}{M}{}-\gss{Z}{M}{-1}$.
\begin{figure*}
\begin{equation}
\gss{\tilde{C}}{}{(t)}\triangleq \diag(\gss{\jim}{M}{*})\gss{C}{}{(t)}\diag(\gss{\jim}{N}{*})=
\gss{D}{}{(t)}-\jim (B_{0}\gss{Z}{M}{-}\gss{D}{}{(t)}\gss{S}{N}{}+\Gamma\gss{D}{}{(t)}\gss{\mathfrak{Z}}{N}{+}-\Im\{\Delta_{0}\}\gss{Z}{M}{-}\gss{D}{}{(t)}\gss{S}{N}{}\gss{\mathfrak{Z}}{N}{-}).
\label{eq:tildeC(X)}
\end{equation}
\hrulefill
\end{figure*}
This provides a way of extracting the (PAM) input symbols from the virtual ones, namely
\begin{equation}
\gss{D}{}{(t)}=\Re\{\gss{\tilde{C}}{}{(t)}\}.
\label{eq:D=Re(C)}
\end{equation}
This, along with the converse step, of translating the input symbols to the virtual ones as in~\pref{eq:tildeC(X)}, will be very helpful in implementing the semi-blind receiver for the FBMC/OQAM case.  

The discussion will henceforth focus on the SIMO setup, to keep it as simple as possible, and to benefit from the guaranteed identifiability of the factors in this case. With $N_{\T}=1$ and hence $\g{\Gamma}=\gss{I}{M}{}$, the sum of Khatri-Rao products structure of $\gss{\bar{Y}}{(1)}{\T}$ described previously reduces to a single Khatri-Rao product, that is, each of the columns of $\gss{Y}{(1)}{\T}$ can be approximated by a Kronecker product. It thus follows from~\pref{eq:Y1eq} that the channel and (virtual) symbol matrices can be determined (up to scaling ambiguity) through a \emph{Khatri-Rao factorization (KRF)}: compute the rank-1 LS approximation (with the aid of a singular value decomposition (SVD)\footnote{An alternative method, of linear instead of cubic complexity, is proposed in~\cite{r15,rrc15} for MIMO-OFDM, however with rather poor results for the channel matrix estimation.}) of the matricized version of each column of $\gss{Y}{(1)}{\T}$~\cite{v15} to come up with estimates (up to scaling ambiguity) of the corresponding columns of $\g{H}$ and $\gss{C}{}{\T}$. Some (limited) training information can be exploited to resolve the scaling ambiguity. 

An alternative solution approach is to exploit the other two modal unfoldings of $\g{\mathcal{Y}}$, namely
\begin{eqnarray}
\gss{Y}{(2)}{\T} & = & \bmx{cccc} \gss{Y}{}{(1)\T} & \gss{Y}{}{(2)\T} & \cdots & \gss{Y}{}{(N_{\mathrm{R}})\T} \emx^{\T} \nonumber \\
& = & (\g{H} \diamond \g{\Gamma})\g{C}+\gss{W}{(2)}{\T}
\label{eq:Y2eq}
\end{eqnarray}
and
\begin{eqnarray}
\gss{Y}{(3)}{\T} & = & \bmx{cccc} \ve(\gss{Y}{}{(1)}) & \ve(\gss{Y}{}{(2)}) & \cdots & \ve(\gss{Y}{}{(N_{\mathrm{R}})}) \emx \nonumber \\
& = & (\gss{C}{}{\mathrm{T}}\diamond\g{\Gamma})\gss{H}{}{\T}+\gss{W}{(3)}{\T},
\label{eq:Y3eq} 
\end{eqnarray}
to alternately solve for $\g{C}$ (given $\g{H}$) and $\g{H}$ (given $\g{C}$), respectively. This results in the classical \emph{alternating least squares (ALS)} algorithm\footnote{The fact that one of the three factor matrices is known could justify a characterization of this problem as a bilinear instead of a trilinear one. To make this explicit, such an ALS algorithm has also been known with the name \emph{bilinear ALS (BALS)}~\cite{rkx12}.} for CPD approximation~\cite{sdfhpf17}, consisting here of iteratively alternating between the conditional\footnote{Also referred to as componentwise optimization~\cite{v15}.} LS updates 
\begin{equation}
\g{\hat{C}}=(\g{H} \diamond \g{\Gamma})^{\dagger}\gss{Y}{(2)}{\T}
\label{eq:Citer}
\end{equation}
and
\begin{equation}
\g{\hat{H}}=\left[(\gss{C}{}{\mathrm{T}}\diamond\g{\Gamma})^{\dagger}\gss{Y}{(3)}{\T}\right]^{\mathrm{T}}.
\label{eq:Hiter} 
\end{equation}
Observe that the permutation ambiguity is trivially resolved in this context because one of the factor matrices, $\g{\Gamma}$, is known, similarly with~\cite{bs02}. A straightforward (and common) way to address the scaling ambiguity is through the transmission of a short training preamble.\footnote{Alternative ways include appropriate normalization of one of the factors (e.g.,~\cite{rkx12}) or the transmission of a pilot sequence at one of the subcarriers (e.g., \cite{js03,bs02}).}
The above updates can be equivalently written (cf.~Appendix~\ref{sec:noise}) as
\begin{equation}
\hat{c}_{m,n}=\frac{\sum_{r=1}^{N_{\mathrm{R}}}H_{m}^{(r)*}y_{m,n}^{(r)}}{\sum_{r=1}^{N_{\mathrm{R}}}|H_{m}^{(r)}|^2}
\label{eq:cmn}
\end{equation}
and
\begin{equation}
\hat{H}_{m}^{(r)}=\frac{\sum_{n=0}^{N-1}c_{m,n}^{*}y_{m,n}}{\sum_{n=0}^{N-1}|c_{m,n}|^2},
\label{eq:Hrm}
\end{equation}
respectively.

\noindent
\emph{Remarks.} 
\begin{enumerate}
\item The well-known expressions for channel equalization,
\begin{equation}
\hat{c}_{m,n}=\frac{1}{N_{\mathrm{R}}}\sum_{r=1}^{N_{\mathrm{R}}}\frac{y^{(r)}_{m,n}}{H_{m}^{(r)}},
\label{eq:IAMc}
\end{equation}
and estimation,
\begin{equation}
\hat{H}_{m}^{(r)}=\frac{1}{N}\sum_{n=0}^{N-1}\frac{y^{(r)}_{m,n}}{c_{m,n}},
\label{eq:IAMH}
\end{equation}
result, respectively, as special cases only for the generally non realistic scenarios of flat magnitude frequency responses and all of the (virtual) symbols having the same magnitude. They consist of solving~\pref{eq:ypqclowMIMO} for $c$ and $H$ in the LS sense, with noise averaging in the spatial and temporal directions, respectively. Using array processing terminology~\cite{kw97}, they are equivalent to \emph{equal-gain combining (ECG)}, whereas the general ALS iterations above correspond to \emph{maximum-ratio combining (MRC)} operations.
\item Interestingly, the above iterations can be also read in recent works on joint estimation for OFDM-~\cite{w13} and FBMC/OQAM-based~\cite{smjv19} MIMO systems. This reveals the hidden tensorial nature of those works and points to the importance of re-visiting iterative Bayesian schemes (e.g., \cite{pmr15}) from this perspective. 
\end{enumerate}

\section{Simulation Results}
\label{sec:sims}

The ALS approach is evaluated in a SIMO $1\times 2$ setup employing two notable instances of flexible multicarrier modulation, namely the conventional CP-OFDM and the maximum spectral efficiency OQAM-based FBMC scheme. The input signal is organized in frames of 53~OFDM (i.e., 106~FBMC/OQAM) symbols each, which are transmitted on $M=32$ subcarriers, using quadrature phase shift keying (QPSK) modulation. CP-OFDM employs a CP of $\frac{M}{4}=8$ samples. Filter banks designed as in~\cite{b01} implement FBMC/OQAM, with $K=4$. PedA channels are considered, which, for a subcarrier spacing of 15~kHz, are of length $L_h=9$ and satisfy the model assumption~\pref{eq:ypqclowMIMO} only very crudely. 
At convergence, and once the complex scaling ambiguity has been resolved, the transmitted symbols can be detected in the CP-OFDM and FBMC/OQAM cases as 
\begin{equation}
\g{\hat{D}}=\mathrm{dec}\left(\diag([e^{\jim\frac{2\pi}{M}mM_{\mathrm{CP}}}]_{m=0}^{M-1})\g{\hat{C}}\odot\gss{\mathfrak{M}}{}{*}\right)
\label{eq:D=dec(C)}
\end{equation}
and
\begin{equation}
\g{\hat{D}}=\mathrm{dec}\left(\Re\left\{\diag(\gss{\jim}{M}{*})\g{\hat{C}}\diag(\gss{\jim}{N}{*})\right\}\right),
\label{eq:D=dec(Re(C))}
\end{equation}
respectively, where $\mathrm{dec}(\cdot)$ signifies the decision device for the input constellation. However, this is `structure blind' in the sense that it does not fully exploit the knowledge of the input constellation and, in FBMC/OQAM, also the information about $\g{D}$ found in the imaginary (interference) part of~\pref{eq:tildeC(X)}. An alternative, `informed' (of the data constellation and interference structure) implementation approach is to include the steps~\pref{eq:D=dec(C)} and \pref{eq:D=dec(Re(C))}, \pref{eq:tildeC(X)} between~\pref{eq:Citer} and~\pref{eq:Hiter} in each ALS iteration. In fact, this corresponds to a projected gradient descent realization of the ALS procedure~\cite{cuc16}. 

\begin{figure}
\begin{center}
\includegraphics[width=7.5cm]{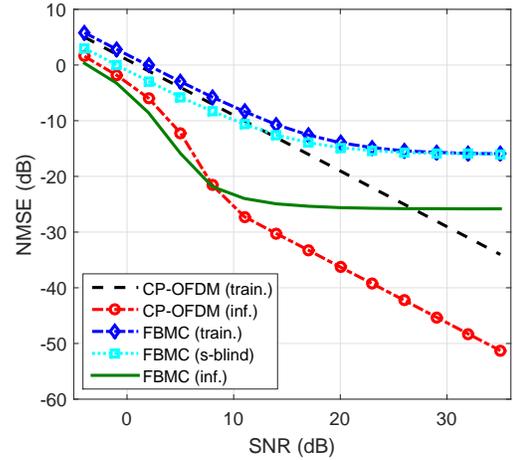} \\
(a) \\
\includegraphics[width=7.5cm]{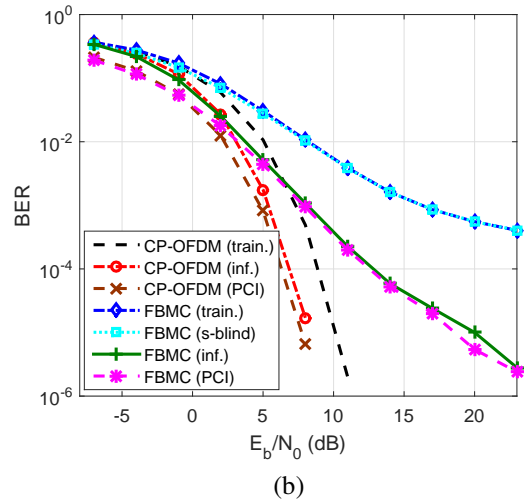} \\
 (b) \\
\includegraphics[width=7.5cm]{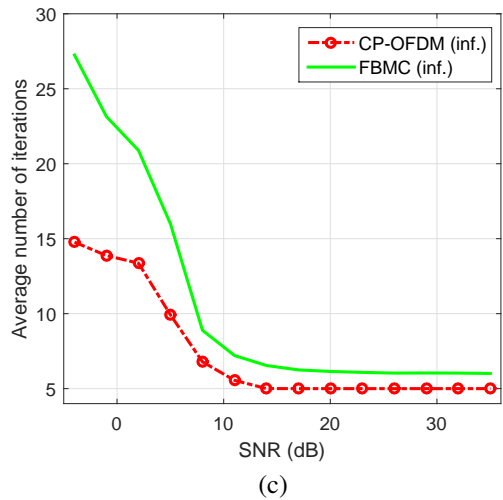} \\
(c) 
\end{center}
  \caption{Performance comparison for a $1\times 2$ system with PedA channels: (a) MSE (b) BER (c) average number of iterations.}
  \label{fig:PedA}
\end{figure}

\begin{figure}
\centering
\includegraphics[width=7.5cm]{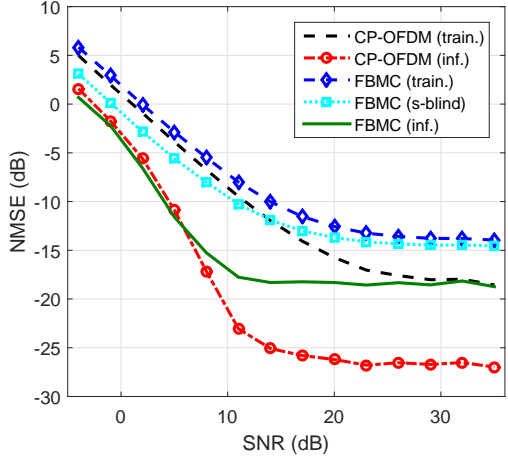} 
  \caption{As in Fig.~\ref{fig:PedA}(a), with VehB channels.} 
  \label{fig:VehB}
\end{figure}

The estimation performance, in terms of normalized mean squared error (NMSE), $E\left\{\left\|\g{H}-\g{\hat{H}}\right\|_{\mathrm{F}}^{2}/\left\|\g{H}\right\|_{\mathrm{F}}^{2}\right\}$, versus transmit signal to noise ratio (SNR), is shown in Fig.~\ref{fig:PedA}(a).
The iterations are initialized with estimates based on MSE-optimal training preambles\footnote{consisting of equipowered pseudo-random pilots for CP-OFDM and the optimal PAM preamble (IAM-R) for FBMC~\cite{kkrt13}.}. Taking in FBMC/OQAM the data constellation and the interference structure into account (see ``inf"(ormed) curves) results in considerable performance gains over the structure-blind approach (``s-blind" curves) and is seen to outperform informed CP-OFDM at low (to medium) SNR values (albeit at the cost of a somewhat larger number of iterations, as seen in Fig.~\ref{fig:PedA}(c)). Moreover, as expected, jointly estimating the channel and the data symbols brings significant improvement over the training only-based approach (``train." curves) since the information about the channel and the symbols underlying the \emph{entire} frame is exploited. Analogous conclusions can be drawn from the bit error rate (BER) detection performance depicted in Fig.~\ref{fig:PedA}(b). Notably, the informed ALS approach is observed to yield results quite close to those obtained when perfect channel information (``PCI") is available. The FBMC/OQAM curves are seen to \emph{floor} at higher SNR values, resulting in performance losses compared to CP-OFDM at such SNR regimes. This is a typical effect of the residual self interference which comes from the invalidation of model~\pref{eq:ypqclowMIMO} and shows up in the absence of strong noise~\cite{kkrt13}. 
These error floors are more severe for more frequency selective channels~\cite{k14}, as shown in the example of VehB channels of length $L_h=18$ depicted in Fig.~\ref{fig:VehB}. 
The CP-OFDM error floors in that case are due to the inadequate length of the employed CP. 

\section{Conclusions}
\label{sec:concls}

The problem of joint channel estimation / data detection (JCD) given limited training input was studied in this paper for the so-called \emph{flexible} MIMO multicarrier system. Several well-known multicarrier modulation schemes, including CP-OFDM and FBMC/OQAM among others, are covered by this large class of modulation waveforms.  A tensor-based approach was followed, which extends earlier related work on MIMO OFDM and FBMC/OQAM systems, and capitalizes on the inherent ability of tensor models to represent multi-dimensional systems and to blindly recover their parameters and inputs. 
Emphasis was put on the SIMO configuration, and on an ALS-type JCD algorithm for determining the channel and symbol matrices, as it encompasses existing JCD approaches as special cases, adding to their understanding and paving the way to further developments. Taking the constellation of the input and the structure of the FBMC self interference into account was demonstrated to offer significant performance gains, in terms of both estimation and detection accuracy, over the non-structure aided and the training only-based approaches, for both mildly and highly frequency selective channels.

Perfect frequency synchronization was assumed (although the analysis did not exclude the presence of a CFO), both for the sake of simplicity and because of the difficulties in CFO estimation that a frequency-domain approach like the one followed in this paper entails in contrast to the time-domain approach taken in other, MIMO-OFDM related works, such as, for example, \cite{js03,bwds17}.\footnote{Working in the frequency domain, of course, has important advantages, especially in a multi-user context~\cite{mtb15}.} As seen in~\pref{eq:ypqclow}, a \emph{time-varying} effective (or global~\cite{laba17}) channel results in the presence of a frequency offset, suggesting the adoption of adaptive techniques (e.g., \cite{laba17}) in general. An effective channel that is still time-invariant results if the CFO factor in~\pref{eq:ypqclow} is approximated as having a fixed $q=L_g/2$ based on the fact that usually both the Tx and Rx pulses peak at that time point~\cite{ssvb18}. 

Future work should aim at taking non-perfect synchronization also into account (including CFO~\cite{glm14} and explicit time offset in a multi-user context~\cite{loz13}) as well as other impairments, such as I/Q imbalance~\cite{w13,cxhym19} and phase noise~\cite{wzj16}.
Further extensions should include channels of strong frequency- (i.e., not satisfying~\pref{eq:ypqclowMIMO}~\cite{kbmbs17}) and time- (low~\cite{rcs06} or high~\cite{cyxll19}) selectivity, as well as richer configurations involving precoders and space-time/frequency coding (e.g., \cite{lcsa13}).

\appendices

\section{Time-Frequency Neighborhood}
\label{sec:TF}

The $I_{u,v}$'s are needed for $u\in \{-1,0,1\}$ and $v\in\{-1,0,1\}$, and obviously $I_{0,0}=1$. For the remaining cases: 

1) $v=0,u=\pm1$

\medskip

\noindent
In this case, \pref{eq:Imnpq} becomes
\[
I_{\pm 1,0} = \varepsilon^{\pm q}\sum_{l=0}^{L_{g}-1}g(l)\tilde{g}(l)e^{\pm \jim \frac{2\pi}{P}l},
\]
and obviously 
\begin{equation}
B_q\triangleq I_{1,0}=I_{-1,0}^{*}=\varepsilon^q \sum_{l=0}^{L_{g}-1}g(l)\tilde{g}(l)e^{\jim \frac{2\pi}{P}l},
\label{eq:Beta}
\end{equation}
where the dependence on $q$ is explicitly shown.
Assuming both $g$ and $\tilde{g}$ are symmetric and of odd length, one can write the above as in~\pref{eq:B},
\begin{figure*}
\begin{equation}
B_q=e^{\jim\frac{2\pi}{P}\left(\frac{L_g-1}{2}\right)}\cdot\varepsilon^q \cdot \left\{r_{0}\left(\frac{L_g-1}{2}\right)+2\sum_{l=0}^{\frac{L_g-1}{2}-1}r_{0}(l)\cos\left[\frac{2\pi}{P}\left(\frac{L_g-1}{2}-l\right)\right]\right\} \label{eq:B}
\end{equation}
\hrulefill
\end{figure*}
where $r_{k}(l)\triangleq g(l+k)\tilde{g}(l)$. Clearly, $B_q$ is a complex-valued quantity in general. If $L_g=KP+1$ (as in, e.g., the PHYDYAS filter~\cite{b01}), the leading exponential factor above becomes $(-1)^K$ and hence $B_0$ is real-valued. If, moreover, $P=M_{\mathrm{ss}}$ or $P=2M_{\mathrm{ss}}$, and hence $\varepsilon=\pm 1$, $B_{q}$ is real for any $q$. For a prototype filter of even length, the above expression still holds by omitting the first term within the braces and changing $\frac{L_g-1}{2}$ to $\frac{L_g}{2}-1$. Obviously, $B_{q+1}=\varepsilon B_q=\varepsilon^{q+1} B_0$.

2) $v=-1,u\in\{-1,0,1\}$

\medskip

\noindent
With $v=-1$, \pref{eq:Imnpq} becomes
\[
I_{u,-1} = \varepsilon^{uq}\sum_{l=0}^{L_{g}-1-M_{\mathrm{ss}}}g(l+M_{\mathrm{ss}})\tilde{g}(l)e^{\jim\frac{2\pi}{P}ul}.
\]
Thus,
\begin{equation}
\Delta_q \triangleq I_{1,-1} = \varepsilon^q\sum_{l=0}^{L_{g}-1-M_{\mathrm{ss}}}r_{M_{\mathrm{ss}}}(l)e^{\jim\frac{2\pi}{P}l}=I_{-1,-1}^{*}.
\label{eq:Delta}
\end{equation}
With $M_{\mathrm{ss}}$ being even as usual and for odd $L_g$ and symmetric $g,\tilde{g}$, an analogous expression as in~\pref{eq:B} holds for $\Delta_q$, namely \pref{eq:D}.
\begin{figure*}
\begin{equation}
\Delta_q=e^{\jim\frac{2\pi}{P}\left(\frac{L_g-M_{\mathrm{ss}}-1}{2}\right)}\cdot \varepsilon^q \cdot
\left\{r_{M_{\mathrm{ss}}}\left(\frac{L_g-M_{\mathrm{ss}}-1}{2}\right)+2\sum_{l=0}^{\frac{L_g-M_{\mathrm{ss}}-1}{2}-1}r_{M_{\mathrm{ss}}}(l)\cos\left[\frac{2\pi}{P}\left(\frac{L_g-M_{\mathrm{ss}}-1}{2}-l\right)\right]\right\} \label{eq:D}
\end{equation}
\hrulefill
\end{figure*}
If $L_g=KP+1$ and $M_{\mathrm{ss}}=\frac{P}{2}$ (as in FBMC/OQAM), the leading exponential factor in \pref{eq:D} becomes $-(-1)^{K}\jim$ and, since $\varepsilon=-1$ in that case, $\Delta_q$ becomes purely imaginary. Clearly, $\Delta_{q+1}=\varepsilon \Delta_q=\varepsilon^{q+1} \Delta_0$. 
For $u=0$, the real quantity
\begin{equation}
\Gamma \triangleq I_{0,-1} = \sum_{l=0}^{L_{g}-1-M_{\mathrm{ss}}}r_{M_{\mathrm{ss}}}(l)
\label{eq:Gamma}
\end{equation}
results and is independent of the time index $q$.

3) $v=1,u\in\{-1,0,1\}$

\medskip

\noindent
Substituting $v=1$ in \pref{eq:Imnpq} yields
\[
I_{u,1} = \varepsilon^{uq}\sum_{l=M_{\mathrm{ss}}}^{L_{g}-1}g(l)g(l-M_{\mathrm{ss}})e^{\jim \frac{2\pi}{P}ul}
\]
and after a change of variable in the summation:
\begin{equation}
I_{u,1} = \varepsilon^{u}\cdot \varepsilon^{uq}\sum_{l=0}^{L_{g}-1-M_{\mathrm{ss}}}r_{M_{\mathrm{ss}}}(l)e^{\jim \frac{2\pi}{P}ul}.
\label{eq:Iu1}
\end{equation}
In view of \pref{eq:Delta},
\begin{equation}
I_{\pm 1,1} = \varepsilon^{\pm 1}I_{\pm 1,-1}.
\label{eq:Delta*}
\end{equation}
Again, the above becomes $I_{\pm 1,1}=I_{\pm 1,-1}$ or $I_{\pm 1,1}=-I_{\pm 1,-1}$ for $P=M_{\mathrm{ss}}$ or $P=2M_{\mathrm{ss}}$, respectively.
Moreover, it follows from \pref{eq:Iu1} and \pref{eq:Gamma} that
\begin{equation}
I_{0,1}=\Gamma.
\label{eq:Gamma*}
\end{equation}

In summary, the $I_{m,n}^{p,q}$ weights in the $3\times 3$ time-frequency neighborhood of the FT point $(p,q)$ are as follows:
\[
\begin{array}{ccc}
\Delta_q^* & B_q^* & (\varepsilon\Delta_q)^* \\
& & \\
\Gamma & 1 & \Gamma \\
& & \\
\Delta_q & B_q & \varepsilon\Delta_q 
\end{array}
\]
or equivalently as in~\pref{eq:pattern}.

\section{More on the Tensor-based Formulation}
\label{sec:PARATUCK2}

The form of the frontal slices of the (noise-free) tensor $\g{\mathcal{\bar{Y}}}$ in~\pref{eq:Yrnu} complies with a PARATUCK-2 decomposition~\cite{fa14}, that is, the combination of a Tucker-2 with a PARAFAC model. Indeed, the tensor follows the Tucker-2 model~\cite{fa14}
\[
\g{\mathcal{\bar{Y}}}=\g{\mathcal{Q}}\times_1 \g{\Gamma} \times_2 \gss{I}{N}{},
\]
where the $N_{\T}M\times N\times N_{\mathrm{R}}$ core tensor $\g{\mathcal{Q}}$ is given by the Hadamard product (along the common modes),
\[
\g{\mathcal{Q}}=\g{C}\odot \g{\mathcal{F}},
\]
of the $N_{\T}M\times N$ matrix $\g{C}$ and the $N_{\T}M\times N\times N_{\mathrm{R}}$ tensor $\g{\mathcal{F}}$ admitting the CPD
\[
\g{\mathcal{F}}=\gss{\mathcal{I}}{3,N_{\mathrm{R}}}{}\times_1 \gss{H}{}{\T} \times_2 \gss{E}{\nu}{},
\]
with $\gss{E}{\nu}{}\triangleq \gss{1}{N_{\mathrm{R}}}{\T}\otimes \gss{e}{\nu}{}$. Following~\cite[eqs.~(70)]{fa14}, one can re-write the above PARATUCK-2 decomposition as an equivalent \emph{constrained} version of~\pref{eq:CPD} as follows:
\begin{equation}
\g{\mathcal{\bar{Y}}}=\gss{\mathcal{I}}{3,NN_{\T}M}{}\times_1 \g{\mathfrak{G}}\times_2 \g{\mathfrak{I}}\times_3 \g{\mathfrak{F}},
\label{eq:Ybar}
\end{equation}
where $\g{\mathfrak{G}}\triangleq \g{\Gamma}(\gss{I}{N_{\T}M}{}\otimes\gss{1}{N}{\T})=\gss{1}{N_{\T}}{\T}\otimes\gss{I}{M}{}\otimes\gss{1}{N}{\T}=\g{\Gamma}\otimes\gss{1}{N}{\T}$, $\g{\mathfrak{I}}\triangleq 
\gss{I}{N}{}(\gss{1}{N_{\T}M}{\T}\otimes\gss{I}{N}{})=\gss{1}{N_{\T}M}{\T}\otimes\gss{I}{N}{}$, and 
\begin{eqnarray*}
\g{\mathfrak{F}} & \triangleq & (\gss{H}{}{\T}\diamond \gss{E}{\nu}{})^{\T}\diag(\ve(\gss{C}{}{\T})) \\
& = & (\underbrace{\gss{H}{}{\T}\otimes \gss{e}{\nu}{}}_{\sbm{H}_{\nu}^{\T}})^{\T}\diag(\ve(\gss{C}{}{\T})).
\end{eqnarray*}
Observe that it is the channel matrix that is combined with the CFO in this expression. An equivalent version of the latter, where $\gss{C}{\nu}{}$ appears instead, is as follows:\footnote{This can also be viewed as a modal unfolding of a CPD with factors $\gss{I}{N_{\T}M}{}$, $\gss{C}{\nu}{\T}$, and $\g{H}$.}
\[
\g{\mathfrak{F}}=\g{H}(\gss{I}{N_{\T}M}{}\diamond \gss{C}{\nu}{\T})^{\T}.
\]
Using this expression in the mode-3 matricization of~\pref{eq:Ybar},
\[
\gss{\bar{Y}}{(3)}{}=\g{\mathfrak{F}}(\g{\mathfrak{J}}\diamond\g{\mathfrak{G}})^{\T},
\]
results, after some algebra, in the expression corresponding to~\pref{eq:CPD}, namely $\gss{\bar{Y}}{(3)}{}=\g{H}(\gss{C}{\nu}{\T}\diamond\g{\Gamma})^{\T}$. The reader is referred to~\cite{k19} for more details on the tensorial formulation of the flexible multicarrier transceiver, including a PARATUCK-2 model for the modulator.  

The combination of symbols and CFO is used in~\cite{bwds17} to (somewhat improperly\footnote{The PARALIND model consists of PARAFAC with \emph{fixed} and \emph{known} linear dependencies on some of its factors, not true in~\cite{bwds17}.}) refer to the tensor model for the MIMO-OFDM system as PARAllel profiles with LINear Dependencies (PARALIND). The way the CFO is determined in~\cite{bwds17} for given channel and symbol matrices would be described in the context of this paper as
\[
\ve(\gss{Y}{(2)}{\T})=(\gss{I}{N}{}\diamond((\g{\hat{H}}\diamond\g{\Gamma})\g{\hat{C}}))\gss{e}{\nu}{}+\ve(\gss{W}{(2)}{\T}),
\]
where \cite[eq.~(27)]{lt08} has been made use of. This amounts to estimating $[\gss{e}{\nu}{}]_{n+1}$ by elementwise dividing (and averaging the result) the corresponding columns of $\gss{Y}{(2)}{\T}$ and $\gss{\hat{Y}}{(2)}{\T}\triangleq (\g{\hat{H}}\diamond\g{\Gamma})\g{\hat{C}}$.

It should be noted that, in contrast to unconstrained PARAFAC, uniqueness conditions for constrained CPD models (such as PARALIND~\cite{sl12}) are themselves also constrained to \emph{partial} (identifiability of some of the factors or part of a factor only) uniqueness~\cite{fa14,cuc16}. 

\section{Channel Estimation and Data Detection with Colored Noise}
\label{sec:noise}

The discussion in this appendix focuses on the SIMO setup, hence $\g{\Gamma}=\gss{I}{M}{}$.
Consider first~\pref{eq:Y3eq} in its equivalent vectorized form,
\begin{equation}
\underbrace{\ve(\gss{Y}{(3)}{\T})}_{\gss{y}{3}{}}=\left[\gss{I}{N_{\mathrm{R}}}{}\otimes (\gss{C}{}{\T}\diamond \boldsymbol{\Gamma})\right]
\underbrace{\ve(\gss{H}{}{\T})}_{\g{h}}+\underbrace{\ve(\gss{W}{(3)}{\T})}_{\gss{w}{3}{}},
\label{eq:Y3vec}
\end{equation}
where use has been made of the well-known property of the vectorized form of a matrix product~\cite{hs81}.  
In view of the assumption that the noise signals at the receiver front ends are identically distributed and uncorrelated to each other, the covariance matrix of the noise vector  
\begin{eqnarray}
\lefteqn{\gss{w}{3}{}} & & \!\!\!\!\!\! \label{eq:w3} \\
& = & \!\!\!\!\!\! \bmx{cccc} \ve(\gss{W}{}{(1)})^{\T} & \ve(\gss{W}{}{(2)})^{\T} & \cdots & \ve(\gss{W}{}{(N_{\mathrm{R}})})^{\T} \emx^{\T} \nonumber \\
& \equiv & \!\!\!\!\!\!\bmx{cccc} \gss{w}{3}{(1)\T} & \gss{w}{3}{(2)\T} & \cdots & \gss{w}{3}{(N_{\mathrm{R}})\T} \emx^{\T} \nonumber 
\end{eqnarray}
will be of the form
\begin{equation}
\gss{C}{\sbm{w}_3}{} \triangleq E\{\gss{w}{3}{}\gss{w}{3}{\He}\}=\gss{I}{N_{\mathrm{R}}}{}\otimes \gss{\bar{C}}{\sbm{w}_3}{},
\label{eq:Cw3}
\end{equation}
where $\gss{\bar{C}}{\sbm{w}_3}{}\triangleq E\{\gss{w}{3}{(r)}\gss{w}{3}{(r)\He}\}$ stands for the noise covariance matrix per receive antenna. 
The block diagonal structure of~\pref{eq:Cw3} implies that the channel responses corresponding to the $N_{\mathrm{R}}$ receive antennas can be estimated separately. To see this in detail, consider the ML estimate as resulting  from~\pref{eq:Y3vec}~\cite{k93}
\begin{eqnarray}
\g{\hat{h}} & = & \bmx{cccc} (\gss{\hat{h}}{}{(1)})^{\T} & (\gss{\hat{h}}{}{(2)})^{\T} & \cdots & (\gss{\hat{h}}{}{(N_{\mathrm{R}})})^{\T} \emx^{\T} \nonumber \\
& = & \left[(\gss{I}{N_{\mathrm{R}}}{}\otimes (\gss{C}{}{\T}\diamond \boldsymbol{\Gamma}))^{\He}\gss{C}{\sbm{w}_3}{-1}(\gss{I}{N_{\mathrm{R}}}{}\otimes (\gss{C}{}{\T}\diamond \boldsymbol{\Gamma}))\right]^{-1}\times \nonumber \\
& & (\gss{I}{N_{\mathrm{R}}}{}\otimes (\gss{C}{}{\T}\diamond \boldsymbol{\Gamma}))^{\He}\gss{C}{\sbm{w}_3}{-1}\gss{y}{3}{}
\label{eq:h1}
\end{eqnarray}
or equivalently, invoking~\pref{eq:Cw3}, \pref{eq:Cbarw3} and properties of the Kronecker product~\cite{hs81}, 
\begin{eqnarray}
\lefteqn{\gss{\hat{h}}{}{(r)}=} & & \label{eq:Hiter_wls} \\
& & \!\!\!\!\!\!\!\!\!\! \left[(\gss{C}{}{\T}\diamond \boldsymbol{\Gamma})^{\He}\gss{\bar{B}}{}{-1}(\gss{C}{}{\T}\diamond \boldsymbol{\Gamma})\right]^{-1}(\gss{C}{}{\T}\diamond \boldsymbol{\Gamma})^{\He}\gss{\bar{B}}{}{-1}\ve(\gss{Y}{}{(r)}), \nonumber
\end{eqnarray}
for the $r$th CFR estimate, $r=1,2,\ldots,N_{\mathrm{R}}$,
with $\g{\bar{B}}\triangleq \gss{\bar{C}}{\sbms{w}{3}}{}/\sigma^2$ denoting the normalized covariance matrix.

In an analogous manner, eq.~\pref{eq:Y2eq} can be written in vectorized form as
\begin{equation}
\underbrace{\ve(\gss{Y}{(2)}{\T})}_{\gss{y}{2}{}}=\left[\gss{I}{N}{}\otimes (\g{H}\diamond \boldsymbol{\Gamma})\right]\underbrace{\ve(\g{C})}_{\g{c}}
+\underbrace{\ve(\gss{W}{(2)}{\T})}_{\gss{w}{2}{}}
\label{eq:Y2vec}
\end{equation}
where
\begin{equation}
\gss{W}{(2)}{}=\bmx{cccc} (\gss{W}{}{(1)})^{\T} & (\gss{W}{}{(2)})^{\T} & \cdots & (\gss{W}{}{(N_{\mathrm{R}})})^{\T}\emx
\label{eq:W2}
\end{equation}
is a re-arrangement of $\gss{W}{(3)}{}$. In fact, it is readily verified that their vectorized versions are related as
\begin{equation}
\gss{w}{2}{}=\gss{P}{2,3}{}\gss{w}{3}{},
\label{eq:w2=Pw3}
\end{equation}
where $\gss{P}{2,3}{}$ is a permutation matrix of order $N_{\mathrm{R}}NM$, given by
\begin{equation}
\gss{P}{2,3}{}=\gss{I}{N_{\mathrm{R}},N}{}\otimes \gss{I}{M}{},
\label{eq:P}
\end{equation}
with $\gss{I}{N_{\mathrm{R}},N}{}$ denoting (in the notation of~\cite{hs81}) the vec-permutation matrix of order $N_{\mathrm{R}}N$.
It then follows that the covariance matrix of $\gss{w}{2}{}$ is~\cite{v05}
\begin{equation}
\gss{C}{\sbm{w}_2}{}=\gss{P}{2,3}{}\gss{C}{\sbm{w}_3}{}\gss{P}{2,3}{\T}
\label{eq:Cw2=PCw3P}
\end{equation}
Using the previous relations in the analogous to~\pref{eq:h1} expression for the ML estimate of $\g{c}$ yields
\begin{eqnarray}
\lefteqn{\g{\hat{c}}=} & & \nonumber \\
& & \left[(\gss{I}{N}{}\otimes (\g{H}\diamond \boldsymbol{\Gamma})^{\He})\gss{P}{2,3}{}\gss{C}{\sbm{w}_3}{-1}\gss{P}{2,3}{\T}
(\gss{I}{N}{}\otimes (\g{H}\diamond\boldsymbol{\Gamma}))\right]^{-1} \times \nonumber \\
& & (\gss{I}{N}{}\otimes (\g{H}\diamond \boldsymbol{\Gamma})^{\He})\gss{P}{2,3}{}\gss{C}{\sbm{w}_3}{-1}\gss{P}{2,3}{\T}\gss{y}{2}{}
\label{eq:c1}
\end{eqnarray}
Note that, similarly with~\pref{eq:w2=Pw3},
\begin{equation}
\gss{P}{2,3}{\T}\gss{y}{2}{}=\gss{y}{3}{}
\label{eq:y2=Py3}
\end{equation}
Furthermore, it is not difficult to see that
\begin{equation}
\gss{P}{2,3}{\T}(\gss{I}{N}{}\otimes (\g{H}\diamond \boldsymbol{\Gamma}))=\bmx{c} \gss{I}{N}{}\otimes \diag(\gss{h}{}{(1)}) \\ \vdots \\ 
\gss{I}{N}{}\otimes \diag(\gss{h}{}{(N_{\mathrm{R}})}) \emx
\label{eq:P[.]}
\end{equation}
Invoking these relations in~\pref{eq:c1} together with~\pref{eq:Cw3} leads to the more explicit expression for the estimate of the symbols vector shown at the top of the next page (eq.~\pref{eq:c}).
\begin{figure*}
\begin{equation}
\g{\hat{c}}=
\left[\sum_{r=1}^{N_{\mathrm{R}}}(\gss{I}{N}{}\otimes \diag(\gss{h}{}{(r)*}))\gss{\bar{B}}{}{-1}(\gss{I}{N}{}\otimes \diag(\gss{h}{}{(r)}))\right]^{-1} 
\sum_{r=1}^{N_{\mathrm{R}}}(\gss{I}{N}{}\otimes \diag(\gss{h}{}{(r)*}))\gss{\bar{B}}{}{-1}\ve(\gss{Y}{}{(r)})
\label{eq:c}
\end{equation}
\hrulefill
\end{figure*}

For the sake of completeness, consider also the covariance of $\gss{w}{1}{}\triangleq\ve(\gss{W}{(1)}{\T})$ in~\pref{eq:Y1eq}. It is easily verified that 
the permutation matrix transforming $\gss{w}{3}{}$ to $\gss{w}{1}{}$ is given by
\begin{equation}
\gss{P}{1,3}{}=\gss{I}{N_{\mathrm{R}}N,M}{},
\label{eq:P13}
\end{equation}
hence
\begin{equation}
\gss{C}{\sbm{w}_1}{}=\gss{P}{1,3}{}\gss{C}{\sbm{w}_3}{}\gss{P}{1,3}{\T}.
\label{eq:Cw1}
\end{equation}

Following~\pref{eq:wpq}:
\[
E\{w^{(r)}_{p,q}(w^{(r)}_{m,n})^*\}=\sigma^2 \tilde{I}_{m,n}^{p,q},
\]
where $\tilde{I}_{m,n}^{p,q}$ is as in~\pref{eq:Imnpq} with $g=\tilde{g}$ and it follows~\pref{eq:pattern} with the corresponding tilded versions of $B_q,\Gamma,\Delta_q$. Hence the noise covariance   
has the following block tridiagonal structure
\begin{eqnarray}
\gss{\bar{C}}{\sbm{w}_3}{} \!\!\!\!\! & = & \!\!\!\!\! \sigma^{2}
\bmx{cccccc} 
\gss{B}{0}{} & \gss{T}{1}{} & \g{0} & \g{0} & \cdots & \g{0} \\ \gss{T}{1}{} & \gss{B}{1}{} & \gss{T}{2}{} & \g{0} & \cdots & \g{0} \\
\g{0} & \gss{T}{2}{} & \gss{B}{2}{} & \gss{T}{3}{} & \cdots & \g{0} \\
\ddots & \ddots & \ddots & \ddots & \ddots & \ddots \\
\g{0} & \cdots & \g{0} & \gss{T}{N-2}{} & \gss{B}{N-2}{} & \gss{T}{N-1}{} \\
\g{0} & \cdots & \g{0} & \g{0} & \gss{T}{N-1}{} & \gss{B}{N-1}{}
\emx \nonumber \\
\!\!\!\!\! & \equiv & \!\!\!\!\! \sigma^{2}\g{\bar{B}},
\label{eq:Cbarw3}
\end{eqnarray}
with the $M\times M$ matrix 
\begin{equation}
\gss{B}{q}{}=\bmx{ccccccc} 1 & \tilde{B}_q & 0 & 0 & \cdots & 0 & \tilde{B}_q \\ \tilde{B}_q^* & 1 & \tilde{B}_q & 0 & \cdots & 0 & 0 \\ 0 & \tilde{B}_q^* & 1 & \tilde{B}_q & \cdots & 0 & 0 \\ \vdots & \ddots & \ddots & \ddots & \ddots & \ddots & \vdots \\ \tilde{B}_q^* & 0 & 0 & 0 & \cdots & \tilde{B}_q^* & 1 \emx
\label{eq:Bmat}
\end{equation}
being the (normalized) covariance matrix of the FBMC noise at time $q$ in the frequency direction, and its (normalized) covariance in the time direction being given by the $M\times M$ matrix
\begin{eqnarray}
\!\!\!\gss{T}{q}{} \!\!\!\!\! & = & \!\!\!\!\!\!\!
\bmx{ccccccc} 
\tilde{\Gamma} & \tilde{\Delta}_q & 0 & 0 & \cdots & 0 & \tilde{\Delta}_q \\ 
\tilde{\Delta}_q^* & \tilde{\Gamma} & \tilde{\Delta}_q & 0 & \cdots & 0 & 0 \\
0 & \tilde{\Delta}_q^* & \tilde{\Gamma} & \tilde{\Delta}_q & \cdots & 0 & 0 \\
\vdots & \ddots & \ddots & \ddots & \vdots & \ddots & \vdots \\
\tilde{\Delta}_q^* & 0 & 0 & 0 & \cdots & \tilde{\Delta}_q^* & \tilde{\Gamma} 
\emx.
\label{eq:T}
\end{eqnarray}
Observe that the matrices in \pref{eq:Bmat}, \pref{eq:T} are banded (tridiagonal), with Hermitian symmetry and Toeplitz 
(almost\footnote{Except for some minor discrepancies at their boundaries, which can be
overlooked for large enough $M$.} circulant) structure. 
For $P=M_{\mathrm{ss}}$, $\varepsilon$ equals~1 and the matrix $\g{\bar{B}}$ above becomes block Toeplitz (and almost\footnote{For $N=2$ or (relatively to $M$) large enough $N$.} block circulant) with Toeplitz (almost circulant) blocks:
\begin{eqnarray}
\g{\bar{B}} \!\!\!\!\! & = & \!\!\!\!\!
\bmx{cccccc} 
\gss{B}{0}{} & \gss{T}{0}{} & \g{0} & \g{0} & \cdots & \g{0} \\ \gss{T}{0}{} & \gss{B}{0}{} & \gss{T}{0}{} & \g{0} & \cdots & \g{0} \\
\g{0} & \gss{T}{0}{} & \gss{B}{0}{} & \gss{T}{0}{} & \cdots & \g{0} \\
\ddots & \ddots & \ddots & \ddots & \ddots & \ddots \\
\g{0} & \cdots & \g{0} & \gss{T}{0}{} & \gss{B}{0}{} & \gss{T}{0}{} \\
\g{0} & \cdots & \g{0} & \g{0} & \gss{T}{0}{} & \gss{B}{0}{}
\emx.
\label{eq:B0}
\end{eqnarray}
In the other most common case, of $P=2M_{\mathrm{ss}}$ (e.g., in FBMC/OQAM), $\varepsilon$ becomes $-1$, and 
\[
\gss{B}{q}{}=\left\{\begin{array}{ll} \gss{B}{0}{}, & \mbox{even\ }q \\ \gss{S}{M}{}\gss{B}{0}{}\gss{S}{M}{}, & \mbox{odd\ }q \end{array}\right.,
\]
and similarly for $\gss{T}{q}{}$, where the $M\times M$ matrix $\gss{S}{M}{}$ is defined as
\begin{equation}
\gss{S}{M}{}\triangleq \diag(1,-1,1,-1,\ldots,1,-1).
\label{eq:S}
\end{equation}
It is readily verified that the $\g{\bar{B}}$ matrix can then be expressed as 
\begin{eqnarray}
\lefteqn{\g{\bar{B}} =\g{\bar{S}} \times} \nonumber \\
& & \!\!\!\!\!\!\!\!\!\!\!\! \underbrace{
\bmx{cccccc} 
\gss{B}{0}{} & \gss{S}{M}{}\gss{T}{0}{} & \g{0} & \g{0} & \cdots & \g{0} \\ 
\gss{T}{0}{}\gss{S}{M}{} & \gss{B}{0}{} & \gss{S}{M}{}\gss{T}{0}{} & \g{0} & \cdots & \g{0} \\
\g{0} & \gss{T}{0}{}\gss{S}{M}{} & \gss{B}{0}{} & \gss{S}{M}{}\gss{T}{0}{} & \cdots & \g{0} \\
\ddots & \ddots & \ddots & \ddots & \ddots & \ddots \\
\g{0} & \cdots & \g{0} & \gss{T}{0}{}\gss{S}{M}{} & \gss{B}{0}{} & \gss{S}{M}{}\gss{T}{0}{} \\
\g{0} & \cdots & \g{0} & \g{0} & \gss{T}{0}{}\gss{S}{M}{} & \gss{B}{0}{}
\emx}_{\sbm{\bar{B}_{S}}} \g{\bar{S}}, \nonumber \\
& & \label{eq:BS}
\end{eqnarray}
where $\g{\bar{S}}\triangleq \diag(\gss{I}{M}{},\gss{S}{M}{},\gss{I}{M}{},\gss{S}{M}{},\ldots,\gss{I}{M}{},\gss{S}{M}{})$ can be viewed as a block extension of~\pref{eq:S}. Observe that the matrix $\gss{\bar{B}}{\sbm{S}}{}$ is Hermitian and block Toeplitz with Toeplitz diagonal blocks. Its off-diagonal blocks have a so-called \emph{alternating circulant} structure (cf.~\cite{k15} and references therein).

It must be emphasized here that, despite the rich structure of the matrix $\g{\bar{B}}$, namely block tridiagonal\footnote{More generally, block banded (if the prototype filter is such that there exists non-negligible correlation beyond the immediate neighbors).}, block Toeplitz, Hermitian, with structured (banded circulant) blocks, its inversion is far from being an easy problem in general and asks for specialized algorithmic solutions, particularly in view of its large scale in practice 
(see, e.g., \cite{cb10} and references therein). Notably, the square-root factorization of the FBMC/OQAM matrix $\gss{\bar{B}}{\sbm{S}}{}$ for the small case of $N=2$ was recently developed in~\cite{k15} and proved to be far from being straightforward. 

Nevertheless, things are much simpler in the case of~\pref{eq:B0}, if $\g{\bar{B}}$ is approximated\footnote{To be precise, transformed via a low-rank circulant transformation~\cite{cb10}.} by a block circulant matrix with circulant blocks. Indeed, one can then employ results from, e.g., \cite{t73} to arrive at the following factorization
\begin{equation}
\g{\bar{B}}=(\gss{F}{N}{}\otimes\gss{F}{M}{})
\bmx{cccc} 
\gss{\Lambda}{0}{} & \g{0} & \cdots & \g{0} \\ 
\g{0} & \gss{\Lambda}{1}{} & \cdots & \g{0} \\ \vdots & \vdots & \ddots & \vdots \\
\g{0} & \g{0} & \cdots & \gss{\Lambda}{N-1}{} \emx 
(\gss{F}{N}{}\otimes\gss{F}{M}{})^{\He},
\label{eq:Bfact}
\end{equation}
where $\gss{F}{Q}{}$ stands for the (normalized) $Q$-point DFT matrix and the matrix $\gss{\Lambda}{n}{}$, $n=0,1,\ldots,N-1$ is $M\times M$ diagonal with diagonal entries
$\lambda_m(\gss{B}{0}{})+2\lambda_m(\gss{T}{0}{})\cos\left(\frac{2\pi}{N}n\right)$, $m=0,1,\ldots,M-1$, with 
$\lambda_m(\gss{B}{0}{})=1+2\tilde{B}_0 \cos\left(\frac{2\pi}{M}m\right)$ and $\lambda_m(\gss{T}{0}{})=\tilde{\Gamma} + 2\Im\{\tilde{\Delta}_0\} \sin\left(\frac{2\pi}{M}m\right)$ for real $\tilde{B}_0$ and imaginary $\tilde{\Delta}_0$, respectively.\footnote{Observe that the $\gss{\Lambda}{n}{}$ matrices enjoy the symmetry $\gss{\Lambda}{n}{}=\gss{\Lambda}{N-1-n}{}$, $n=0,1,\ldots,N-1$. Moreover, $\lambda_{m}(\gss{B}{0}{})=\lambda_{M-m}(\gss{B}{0}{})$, $m=1,2,\ldots,M-1$, and each of the sequences $\lambda_{m}(\gss{T}{0}{})$, $m=1,2,\ldots,\frac{M}{2}-1$ and $m=\frac{M}{2}+1,\ldots,M-1$ enjoys even symmetry, with $\lambda_0(\gss{T}{0}{})=\lambda_{\frac{M}{2}}(\gss{T}{0}{})$.}
Substituting in~\pref{eq:Hiter_wls} and using well-known properties of the Kronecker product and the $\ve$ operator~\cite{hs81} yields the following expression for the CFR estimate of the $r$th channel:
\begin{eqnarray*}
\lefteqn{\gss{\hat{h}}{}{(r)}=} \\
& & \left[\sum_{n=0}^{N-1}\diag(\gss{\breve{c}}{n}{*})\gss{F}{M}{}\gss{\Lambda}{n}{-1}\gss{F}{M}{\He}\diag(\gss{\breve{c}}{n}{})\right]^{-1}\times \\
& & \sum_{n=0}^{N-1}\diag(\gss{\breve{c}}{n}{*})\gss{F}{M}{}\gss{\Lambda}{n}{-1}\gss{\breve{y}}{n}{(r)},
\end{eqnarray*}
where $\gss{\breve{c}}{n}{\T}$ is the $(n+1)$st row of $\gss{F}{N}{\He}\gss{C}{}{\T}$ and $\gss{\breve{y}}{n}{(r)}$ is the $(n+1)$st column of the IDFT of $\gss{Y}{}{(r)}$ in both its directions, namely $\gss{F}{M}{\He}\gss{Y}{}{(r)}\gss{F}{N}{*}$. It is evident from~\pref{eq:Yr} that the above expression still holds with no IDFT in the time direction, hence 
\begin{eqnarray}
\lefteqn{\gss{\hat{h}}{}{(r)}=} \nonumber \\
& & \left[\sum_{n=0}^{N-1}\diag(\gss{c}{n}{*})\gss{F}{M}{}\gss{\Lambda}{n}{-1}\gss{F}{M}{\He}\diag(\gss{c}{n}{})\right]^{-1}\times \nonumber \\
& & \sum_{n=0}^{N-1}\diag(\gss{c}{n}{*})\gss{F}{M}{}\gss{\Lambda}{n}{-1}\gss{F}{M}{\He}\gss{y}{n}{(r)},
\label{eq:Hr}
\end{eqnarray}
where $\gss{c}{n}{}$ and $\gss{y}{n}{(r)}$ denote the $(n+1)$st columns of $\g{C}$ and $\gss{Y}{}{(r)}$, respectively. 
Substituting for $\g{\bar{B}}$ from~\pref{eq:Bfact} in~\pref{eq:c} and after some algebra results in the expression~\pref{eq:ccolor} for the estimate of the symbol vector. 
\begin{figure*}
\begin{eqnarray}
\g{\hat{c}} & = & 
\left \{\sum_{r=1}^{N_{\mathrm{R}}}(\gss{I}{N}{}\otimes\diag(\gss{h}{}{(r)*})\gss{F}{M}{})
\left[\sum_{n=0}^{N-1}(\gssa{F}{N}{}{:,n+1}\gssa{F}{N}{}{:,n+1}^{\He}\otimes\gss{\Lambda}{n}{-1})\right](\gss{I}{N}{}\otimes\diag(\gss{h}{}{(r)*})\gss{F}{M}{})^{\He}\right\}^{-1} \times \nonumber \\
& &\sum_{r=1}^{N_{\mathrm{R}}}(\gss{I}{N}{}\otimes\diag(\gss{h}{}{(r)*})\gss{F}{M}{})
\sum_{n=0}^{N-1} (\gssa{F}{N}{}{:,n+1}\otimes \gss{\Lambda}{n}{})\gss{\breve{y}}{n}{(r)}.
\label{eq:ccolor}
\end{eqnarray}
\hrulefill
\end{figure*}

Neglecting the noise correlation in~\pref{eq:Hr}, i.e., setting $\g{\bar{B}}=\gss{I}{MN}{}$ (and hence $\gss{\Lambda}{n}{}=\gss{I}{M}{}$, $n=0,1,\ldots,N-1$) above, yields
\[
\gss{\hat{h}}{}{(r)}=
\left[\sum_{n=0}^{N-1}\diag(\gss{c}{n}{*}\odot\gss{c}{n}{})\right]^{-1}
\sum_{n=0}^{N-1}\diag(\gss{c}{n}{*})\gss{y}{n}{(r)},
\]
or equivalently \pref{eq:Hrm}.
In that case, \pref{eq:ccolor} can be seen to become
\[
\g{\hat{C}}=\left[\sum_{r=1}^{N_{\mathrm{R}}}\diag(\gss{h}{}{(r)*}\odot\gss{h}{}{(r)})\right]^{-1}
\sum_{r=1}^{N_{\mathrm{R}}}\diag(\gss{h}{}{(r)*})\gss{Y}{}{(r)},
\]
yielding \pref{eq:cmn}.

\bibliographystyle{IEEEbib}
\bibliography{IEEEabrv,refs}

\end{document}